\documentclass[11pt,a4paper]{article}
\usepackage{jheppub}
\usepackage{booktabs}
\usepackage{subfigure}
\usepackage{multirow}
\usepackage{subfigure}
\usepackage{graphicx}
\usepackage{epsfig}
\usepackage{braket}
\usepackage{lscape}
\usepackage{rotating}

\newcommand{\x}{$\times$}

\title{Framework for Model Independent Analyses \\ of Multiple Extra Quark Scenarios}

\author[a,b]{Daniele Barducci}
\author[a,b]{, Alexander Belyaev}
\author[c]{, Mathieu Buchkremer}
\author[d,e]{, Giacomo Cacciapaglia}
\author[d,e]{, Aldo Deandrea}
\author[f]{, Stefania De Curtis}
\author[g]{, Jad Marrouche}
\author[a,b]{, Stefano Moretti}
\author[a,b]{and Luca Panizzi}

\affiliation[a]{School of Physics and Astronomy, University of Southampton, Highfield, Southampton SO17 1BJ, UK}
\affiliation[b]{Particle Physics Department, Rutherford Appleton Laboratory, Chilton, Didcot, Oxon OX11 0QX, UK}
\affiliation[c]{Centre for Cosmology, Particle Physics and Phenomenology (CP3), Universit\'e catholique de Louvain, Chemin du Cyclotron, 2, B-1348, Louvain-la-Neuve, Belgium}
\affiliation[d]{Universit\'e de Lyon, F-69622 Lyon, France; Universit\'e Lyon 1, Villeurbanne}
\affiliation[e]{CNRS/IN2P3, UMR5822, Institut de Physique Nucl\'eaire de Lyon, F-69622 Villeurbanne Cedex, France}
\affiliation[f]{INFN, Sezione di Firenze, Via G. Sansone 1, 50019 Sesto Fiorentino, Italy}
\affiliation[g]{Physics Department, CERN, CH-1211, Geneva 23, Switzerland}

\abstract{In this paper we present an analysis strategy and a dedicated tool to determine the exclusion confidence level for any scenario involving multiple heavy extra quarks with generic decay channels, as predicted in several extensions of the Standard Model. 
We have created, validated and used a software package, called  XQCAT (eXtra Quark Combined Analysis Tool), which is based on publicly available experimental data from direct searches for top partners and from Supersymmetry inspired searches.
By means of this code, we recast
the limits from CMS on new heavy extra quarks considering a complete set of decay channels. The
resulting exclusion confidence levels are presented for some simple
scenarios with multiple states and general coupling assumptions. Highlighting the importance of
combining multiple topology searches to obtain accurate re-interpretations of the existing
searches, we discuss the reach of the SUSY analyses so as to set bounds on new quark
resonances.  In particular, we report on the re-interpretation of the
existing limits on benchmark scenarios with one and multiple pair-produced top partners
having non-exclusive couplings to the third Standard Model generation of quarks.}

\keywords{Heavy extra quarks, vector-like quarks, LHC, model-independent analysis of experimental data}

\preprint{CP3-14-19, LYCEN 2014-04}

\makeindex

\begin{document}

\maketitle
\flushbottom

\tableofcontents

\section{Introduction}

Since the start of its physics programme, the Large Hadron Collider (LHC) has delivered very high quality results on the discovery of the Higgs boson and the first characterisation of its properties as well as a large set of new limits on searches 
for other particles that  may be present in extensions of the Standard Model (SM). In the next planned runs, the LHC will continue improving 
existing limits on new particles beyond those predicted by the SM or else provide evidence of these new 
states. In this spirit, new physics models should be built by considering both the theoretical ideas and the experimental scope, with the aim of making the former testable via the latter.
 
To establish such a direct link between experimental and theoretical particle physics is not always an easy task though. 
In fact, experimental analyses are performed in terms of final states observed in the detector (particle signatures) and interpreted in the analyses in terms of specific models or classes of models, typically with simplifying assumptions. 
Usually, starting from these existing analyses, more general conclusions on different classes of theoretical models cannot be 
obtained directly. It is indeed impossible to cover any possible theoretical model and the experimental analyses 
are therefore performed either for the most popular specific scenarios or more general 
ones\cite{Alves:2011wf}, like Supersymmetry (SUSY). Conversely, we have seen in recent years that the extra knowledge brought in by the recent collider data is shifting the 
focus of the particle theory community from established Beyond the Standard Model (BSM) scenarios of which we have no evidence (like SUSY) to new ones which
may be promptly testable by experiment. 

In order to follow these changing expectations
for physics beyond the SM, it is interesting to see to what extent ``data recasting" of existing analyses is possible, that is, to re-interpret data selected with a specific BSM scenario in mind into others. In this connection, it will be intriguing to see that data used in SUSY searches lend themselves to perform this
exercise. 

For cut-based analyses a Monte Carlo (MC)  simulation and a simplified detector emulation can allow the 
conversion of existing searches into a set of new cases, but this is typically done on the basis of a single model or a single final state. 
Relying on Monte Carlo (MC) simulations and simplified detector emulations for cut-based analyses, various collaborations have already attempted to tackle this
problem on more general grounds, such as the two
tools
CheckMATE \cite{Drees:2013wra},  
SModelS \cite{Kraml:2013mwa} or Fastlim \cite{Papucci:2014rja}. 
Each code has its own specific way to prepare the input model spectrum (i.e., masses and couplings)
and address the comparison with existing data analyses, the former
adopting a rather generic model independent approach while the others optimising performance for SUSY scenarios, yet in 
all cases the 
full spectrum of a BSM realisation is tested. Our approach is somewhat complementary to both, as it concentrates on one specific `sector' (heavy extra quarks), as opposed to the full model, which may belong to a variety of BSM
frameworks and is promptly and easily testable in hadro-production at the LHC, thereby rendering the hypotheses formulated amenable to prompt experimental confirmation or disproval. In a sense, we sacrifice generality for effectiveness.

In particular, in this paper, we focus on building a general framework and a software tool for 
reinterpreting existing analyses to the case of more generic and realistic scenarios containing multiple heavy eXtra
Quarks (XQs).
The motivation to concentrate on these states is due to the recent renewed interest for 
the presence of vector-like quarks (coloured fermions whose mass is not generated by the Brout-Englert-Higgs mechanism), following, on the one hand, the exclusion of a fourth SM-like family of (chiral) quarks\footnote{However, this exclusion assumes the presence of a single Higgs that has SM couplings to the fourth family and the top quark, and the constraints can be evaded in non-minimal extensions of the SM, see for example \cite{Holdom:2014bla}.} and, on the other hand, the fact that such states are present in many extensions of the SM that offer alternative explanations for the dynamics of
Electro-Weak Symmetry Breaking (EWSB) to the one embedded in the Brout-Englert-Higgs mechanism. In  fact,
such new fermionic resonances, sometimes referred to as ``top partners'' when associated to the top quark 
may also play a crucial role to soften the quadratic divergences contributing to the Higgs mass term, the origin of the SM hierarchy problem. 
Our approach, nevertheless, can be applied both to chiral and vector-like heavy extra quarks, so that we will employ XQ as a common denomination.
New heavy XQs then appear in models of extra dimensions where they arise as Kaluza-Klein excitations of the standard 
fermions \cite{Antoniadis:1990ew,Csaki:2003sh,Cacciapaglia:2009pa}, in models aiming at generating fermion masses\cite{Grossman:1999ra} and in models with extended gauge symmetries, like Little Higgs models\cite{ArkaniHamed:2002qx}. Other examples include scenarios based on partial compositeness \cite{Kaplan:1983fs,Kaplan:1991dc}, 
where the Yukawa couplings of the elementary quarks emerge from their mixings with new heavy composite states, thereby
providing practicable alternative avenues to solving the problem of fermion mass generation. One can then include models with a pseudo-Nambu-Goldstone-Higgs\cite{Agashe:2004rs,Contino:2006qr}, general Composite Higgs Models (CHMs)\cite{Giudice:2007fh}, models of gauge-Higgs\cite{Hosotani:1983xw} and gauge coupling unification (see, e.g., \cite{Choudhury:2001hs,Panico:2008bx}) or extended custodial symmetry protecting the $Zb\bar{b}$ coupling from large deviations \cite{Agashe:2006at,Chivukula:2011jh}. 
After the discovery of the Higgs, most of the attention has been dedicated to XQs 
mixing exclusively with the top quark, in particular in the general scenario of a pseudo-Nambu-Goldstone (or Composite) Higgs state \cite{Anastasiou:2009rv,Marzocca:2012zn,DeSimone:2012fs}. Finally, XQs coupled to the third generation of quarks have been thoroughly investigated from a phenomenological point of view \cite{delAguila:2000rc,AguilarSaavedra:2005pv,AguilarSaavedra:2009es,Cacciapaglia:2010vn,Ellis:2014dza}, together with the possibility to have sizeable couplings to the light generations\cite{Atre:2011ae,Cacciapaglia:2011fx}.

The key point to appreciate, though, is that 
most of such new physics models contain in general more than one new heavy XQ state. Up till now, this has gone against the approach of experimental searches for vector-like quarks, which are undertaken under the simplifying assumption
that only one new XQ state beside the SM spectrum is present and bounds are given on this sole object (on its mass and/or couplings). These bounds cannot be easily re-interpreted in models containing two or more new quarks, which can also contribute to the signal events in the very same channel. Further, different XQs can decay into the same final state through
different decay chains and therefore with different topologies. For either of these reasons, experimental efficiencies would therefore be different and a rescaling of mass bounds is not trivial. Finally, unknown input parameters in the new physics Lagrangian usually affect both the spectrum of (multiple) XQs and their branching ratios (BRs). We are quick to point out, however, that progress on this front is currently underway in the experimental community, in both designing new searches and interpreting these in the context of models where the XQ couplings are either to heavy or light flavour quarks as well as to a linear combination of the two. 
In this work we are therefore trying to approach the question: given the parameter space describing the extra quark sector in a BSM model and the experimental data from searches of new XQs (or indeed others),
would it be possible to find allowed or
excluded parameter regions without performing dedicated simulations? In order to obtain an answer to this question two major issues should be
solved first: the creation of a pre-loaded data-base of existing experimental analyses to be used for this task and a large amount of MC data. These two issues are the most time-consuming 
part for such analyses and the focus of this paper is precisely to provide a way to optimise such tasks by means of  an automated tool in order to considerably 
simplify the comparison. We therefore built a code, named \verb|XQCAT|
\footnote{The code is available at the website \href{https://launchpad.net/xqcat}{https://launchpad.net/xqcat}.} (eXtra Quark Combined Analysis Tool), that allows the user to quickly determine the exclusion 
Confidence Level (eCL\footnote{Defined as $eCL\equiv 1-CL_s$, where the $CL_s$ statistical quantity, and related procedure, is described in \cite{Read:2002hq,Read:2000ru}.}) of any given scenario with multiple
XQs: a scenario is basically defined by providing 
the masses and BRs of each new quark of a given species.
Model independence is achieved considering only pair production, which is induced by 
QCD interactions,
so that the emerging cross section is only sensitive to the masses of the new quarks. A similar procedure can also be applied to single-production channels, as shown in~\cite{Buchkremer:2013bha}, but it has not been included in the current version of the code. It is also assumed that none of the 
new particles decay into each other, which is a 
reasonable expectation for members of a vector-like multiplet with a common mass scale $m$ and small mass splittings, as in most of the models we mentioned above. 
Under these assumptions, the cross sections of various final states can be decomposed in model-independent subsets which contain all the kinematic information 
from the decays. Thanks to this observation, given the information on the masses and BRs, it is finally possible to reconstruct the signal coming from general
scenarios by combining, with the appropriate ``weights'', the different model-independent topologies which generate the signal and the different kinematic
distributions. 
The efficiencies of the various signal channels for the many included searches are encoded in the database of  \verb|XQCAT|, so that no additional MC generation is needed in order to calculate the eCLs.
The signal channels are characterised by the final states and a set of masses of the XQs.
All in all, from a basic input of the code consisting of only masses and BRs of all new XQs, the tool provides the eCL of the specific spectrum combining data from different experimental searches. 
Given the publicly available results at this time, in this preliminary study we chose to use one published CMS analysis\footnote{Other CMS searches as well as ATLAS ones will be included in a forthcoming version of the tool.} based on the entire 8 TeV dataset optimised for a single vector-like XQ coupled to third-generation quarks\cite{Chatrchyan:2013uxa}. By 
definition, one 
would 
expect that this search is not suited to a re-interpretation when considering coupling to light-flavour quarks and so, in addition, we 
use a set of four inclusive topological SUSY searches\cite{Chatrchyan:2012wa,Chatrchyan:2012sca,Chatrchyan:2012te,Chatrchyan:2012sa,Chatrchyan:2013lya,Chatrchyan:2012paa} whose generic jets + 0,1,2 lepton signature should be sensitive to final states with light quarks. The results in this paper are therefore a realistic estimate on the weakest limits that could be achieved on multiple XQ scenarios while new experimental searches and interpretations are in progress.

The paper is organised as follows. In Section 2, we describe the analysis strategy for the \verb|XQCAT|\ code, including the generation of the database of efficiencies
and the calculation of the eCLs. In Section 3 the validation of the code is discussed in detail and tests are thoroughly illustrated 
in simple cases. In Section 4, 
we use the \verb|XQCAT|\ code to provide new constraints on scenarios with multiple XQs. In Section 5, the interplay and complementarity {between direct XQ and SUSY searches is discussed. Section 6 contains the conclusions and emphasises the prospects of applying the code to other existing searches and other new physics scenarios.

\section{Analysis strategy}

\subsection{General approach}

In our model-independent approach, the analysis of scenarios with multiple XQ states that can decay into {\it any} SM quark together with a Higgs or gauge boson is addressed by a dedicated computing tool called \verb|XQCAT|. 

Starting from the QCD pair production followed by the decays of the new heavy quarks,
allowed to mix with 
\textit{any} SM quarks via Yukawa interactions,
one can estimate  the number of signal events that survive the selection cuts 
for  any given signature and specific search strategy, that is finding the respective  efficiencies for  each subprocess that contributes to 
the given final state.
We have simulated  signals for each subprocess  for different values of the masses of the XQs and have then used an analysis code to 
evaluate the respective efficiencies for each subprocess
contributing to each individual signature {and the respective search}.
These efficiencies
are then stored in the database of \verb|XQCAT|
The most general scenario we consider covers the case of the following XQ species which can be simultaneously present in the generic model:
\begin{equation}
n_X X_{5/3} + n_T T_{2/3} + n_B B_{-1/3} + n_Y Y_{-4/3}
\end{equation}
where the sub-script indicates the electric charge.
Therefore, $T$ and $B$ are 
up- and down-like quark partners, respectively, and $X$ and $Y$
are exotic particles with charge 5/3 and $-4/3$, respectively. 
We limit 
our analysis to these four species of XQs because they are the only ones which can couple directly to SM quarks.

The new states are then assumed to decay promptly, otherwise their signatures would be that of long-lived heavy objects or of bound states thereof, depending on their lifetime \cite{Buchkremer:2012dn}. 
Assuming general mixings between XQs and SM quarks, we consider the following complete set of decays:
\begin{itemize}
\item $X \to W^+ u_i$,

\item $T \to W^+ d_i\,, \; Z u_i\,, \;  H u_i$,

\item $B \to W^- u_i\,, \; Z d_i\,, \; H d_i$,

\item $Y \to W^- d_i$,
\end{itemize}
where $i=1,2,3$ labels the SM family. 
Here, we do not consider decays into another XQ plus a SM boson: such decays may be significant only in the case of large mass splitting. Even in this case though, the phenomenology is usually dominated by the lightest state(s) in the model, providing the above decay patterns. Furthermore, additional   decay chains introduce model-dependence and will only strengthen the 
bounds given that they would increase the inclusive production cross section of the lightest XQs.
We also do not consider decays into other non-SM final states: for instance, in some models top and bottom partners can decay into SM quarks and a stable neutral particle that represents a Dark Matter (DM) candidate. Thus, decays of the type $T \to A_{DM} u_i$ and $B \to A_{DM} d_i$ have not been considered yet\cite{Perelstein:2011ds} as they belong to a different class of models: they  will only be taken into account in a forthcoming version  of the code.
Nevertheless, to remain general, the code accept as input BRs that do not add up to 1, i.e. the sum can be smaller than 1 to allow other decays besides those considered above, of course with the caveat that decay modes not listed above will not be included in the calculation of the bounds. 

To illustrate in practice how our strategy works, let us consider a simple example with just one $B$ quark and assume that it
decays only to $Wu$ and $Wt$. Following the  QCD  $B \bar{B}$ pair production these decays  would lead to the following signatures:
\begin{eqnarray*}
p p \to B \bar{B} \to \left\{ \begin{array}{l}
W^+ W^- u \bar u \\ 
W^+ W^- u \bar t \to W^+ W^- W^- u \bar{b}\\ 
W^+ W^- t \bar u \to W^+ W^+ W^- b \bar{u}\\ 
W^+ W^- t \bar t \to W^+ W^+ W^- W^- b \bar b
\end{array} \right.
\end{eqnarray*}
which we will denote as:
$W^+ W^- j j$, $W^+ W^- W^- j \bar{b}$, $W^+ W^+ W^- j b$ and $W^+ W^+ W^- W^- b \bar{b}$. 
In each specific model, the weight of events coming from each channel will be determined by the BRs: $BR_{Wu}$ and $BR_{Wt}$.
If we distinguish channels according to $W$-boson multiplicity as
$WW$, $WWW$ and $WWWW$ channels, then their relative rates are given by
BR$_{Wu}^2$~:~(2\,BR$_{Wu}{\rm  BR}_{Wt})$~:~BR$_{Wt}^2$ respectively.
In addition to different rates, these three different channels have different kinematics for the final state fermions (after the $W$-bosons decays) and in turn different efficiencies  upon the application of the analysis cuts.
This complication is taken into account in our code, in which
we have considered each channel independently and derived the respective
efficiencies as a function of the XQ mass.
Thus, considering  the various BRs of  XQ  as independent 
parameters and using the efficiencies mentioned above, one can evaluate
the overall signal rate as a weighted sum of all channels under study.

This approach can be easily extended to  
 models with more than one XQ: for instance, we can consider a scenario with 
 one $X$,  two $T_{1,2}$ and one $B$ states, each with different masses $m_X$, $m_{T_1}$, $m_{T_2}$ and $m_{B}$ and (in general) BRs. 
This setup can arise naturally in specific models such as ~\cite{DeSimone:2012fs}, as will be considered explicitly in Sect.\ref{sec:MultiXQ}. In this case, two or more different XQs can contribute to the same final state signature.
For example,  if one would like to study
 the sensitivity of an experimental search for events with same-sign
di-leptons, more than 2 jets and more than 2 $b$-jets, then the following channels can contribute to the above final state\cite{Matsedonskyi:2012ym}: $p p \to B \bar{B} \to W^+ \bar{t} W^- t$ and $p p \to X \bar{X} \to W^- \bar{t} W^+ t$.
In such a case, our tool weights the efficiencies of different
 channels by {the different cross sections} and BRs (that depend on the masses $m_X$ and $m_B$), 
finally providing an eCL for the combined signal.

As intimated, for XQ production mode we have only considered QCD pair production: the advantage of this approach is that it is fully model independent, as the cross sections only depend on the masses of the XQs and not on their EW couplings.
Sub-leading pair production channels are provided by EW  processes, however, their cross sections are much smaller than the QCD ones and thus can be safely neglected. In fact, 
as no experimental search targeted at EW pair production is available, including their effect coud only give a minor improvement in our bounds.
Another interesting production channel is given by single production of XQs\cite{Atre:2011ae,Buchkremer:2013bha}.
It is well understood that single production can be relevant, especially for large masses, as it is less suppressed by the
phase space of the final state and Parton
Distribution Functions (PDFs) of the proton.
Furthermore, large cross sections can be expected even after imposing flavour and EW precision bounds on the mixing parameters. The inclusion of single production to our tool strategy is straightforward, e.g., by following \cite{Buchkremer:2013bha}, and  will be implemented  in the near future.

\subsection{Generation of the efficiency database}

The simulation of QCD pair production for each quark and for each mass has
been performed with MadGraph5, v.1.5.8~\cite{Alwall:2011uj}. The subsequent decays of the XQs into SM
quarks and bosons have been computed with BRIDGE, v.2.24~\cite{Meade:2007js}. PYTHIA, v.6.4~\cite{Sjostrand:2006za} has been used for
decays of SM particles and subsequent hadronization and parton shower. The
detector simulation has been performed by Delphes2, v.2.0.2~\cite{Ovyn:2009tx} with
suitable detector card for CMS (analogous card for ATLAS will be used in a next upgrade of the code). Since the XQ pair production
is a QCD process, jet-matching has also been considered and appropriate matching
parameters have been chosen for each mass of the XQs. Therefore, the
processes simulated in MadGraph5 have been: 
\begin{eqnarray}
p p \to Q \bar Q + \{0,1,2\} \mbox{ jets}
\end{eqnarray}
where $Q = T, B, X, Y$.

The number of simulated processes is related to the number of considered
masses, to the possible decay channels and to the chirality of the couplings%
\footnote{%
It is possible to prove~\cite{Buchkremer:2013bha} that the couplings of
vector-like XQs to SM quarks and bosons are dominantly chiral and that the dominant chirality
depends on the representation under $SU(2)$ the XQ belongs to. In any case, the tool assumes a dominant chirality for the XQ couplings, whether they are vector-like or not.}. Each $T$
and $B$ can decay into 9 channels, corresponding to combinations of three SM
bosons and three SM quarks, however, since light generation quarks cannot be distinguished at the LHC and are
experimentally seen as jets, the effective observable decays are just 6 for
both $T$ and $B$ (assuming the bottom quark to be tagged, while we are not considering yet the possibility to tag a charm quark). 
Since we are dealing with QCD pair production, the total number of
combinations is $6\times 6 =36$ for both $T$ and $B$. Exotic XQs, $X$ and $Y$, can only
decay through charged currents, and therefore the possible combinations for
pair production are $2\times 2=4$ for both $X$ and $Y$. Considering two chiralities for
each combination, the total number of channels for each mass scale is 
equal to $2\times 2\times(6\times 6 + 2\times 2)=160$.
The simulation has been performed considering masses in the range $%
\{400,2000\}$ GeV with steps of 100 GeV. 

The calculation of the search efficiencies is performed in a framework built on the Delphes~\cite{Ovyn:2009tx} detector simulation as an input. This framework has been validated and used previously\cite{Buchmueller:2012hv,Buchmueller:2013exa} and is described in\cite{Buchmueller:2013exa}. 

The whole analysis is based on a cut-and-count approach; though more refined techniques are adopted by experimental collaborations, as shape analyses or BDT techniques, it is not possible to accurately reproduce these approaches with the tools at our disposal, as attempting to reproduce the shapes using fast simulation tools such as Delphes introduces large uncertainties on the shapes, which weaken the limits considerably. The cut-and-count approach is therefore the most accurate technique we can adopt with fast simulation tools, and the validation will be performed selecting subsets of bins that reproduce as closely as possible the experimental results.

\bigskip

For the purpose of our present analysis, we have implemented two different kind of searches (limiting to CMS studies for the moment, ATLAS searches will be included in the following version of the code):
\begin{itemize}
\item \underbar{\textit{Direct search of XQs}} We implemented the CMS analysis B2G-12-015 \cite{Chatrchyan:2013uxa}, at $\sqrt{s}=8$ TeV with a 19.5 fb$^{-1}$ dataset, for a pair produced $T$ quark that mixes only with third-generation SM quarks and can decay to $W^+b, Zt$ or $Ht$ with variable BRs. The CMS collaboration presents the 95\% CL lower limits on the $T$ quark mass for different combinations of its BRs using six mutually exclusive channels: two single lepton (single electron and single muon), three di-lepton (two opposite-sign and one same-sign) and one tri-lepton channel, all containing tagged $b$-jets in the final state. No deviations from the SM expectations were observed when considering a large number of benchmark points among the allowed parameter space for the  BRs. Since the sensitivity of the search is mostly driven by the multi-lepton channels, in the present version of the tool we have only implemented three bins:
the opposite-sign 1 (OS1, in the search), the same-sign (SS) and the tri-lepton channels. More details about this choice will be explained in the validation section below. The limits for the multi-lepton channels only can be found in the twiki page of the search \cite{twikiB2G12015} and the quoted observed bounds are in the range $592\div794$ GeV depending on the assumed BRs.
\item \underbar{\textit{SUSY searches}} We implemented four searches inspired by SUSY scenarios, characterised by the presence of different numbers of leptons in the final state and large missing transverse energy: 0-lepton ($\alpha_{T}$) \cite{Chatrchyan:2012wa}, mono-lepton ($L_{p}$) \cite{Chatrchyan:2012sca}, opposite-sign dilepton (OS) \cite{Chatrchyan:2012te} plus same-sign dilepton (SS) \cite{Chatrchyan:2012sa}, considering the entire 4.98 fb$^{-1}$ 2011 dataset at $\sqrt{s}=7$ TeV. We have also included the updated $\alpha _{T}$ \cite{Chatrchyan:2013lya} and same-sign \cite{Chatrchyan:2012paa} searches at 8 TeV, with 11.7 fb$^{-1}$ and 10.5 fb$^{-1}$, respectively. It has been verified that the selected searches are uncorrelated and, therefore, it is possible to statistically combine them, yielding 95\% CL bounds at 7 TeV (combination of 4 searches), 8 TeV (combination of 2 searches) and 7+8 TeV (combination of 6 searches). A validation of the uncorrelation of the SUSY searches has been performed in \
\cite{Buchmueller:2013exa} and relies on the following facts:
\begin{itemize}
 \item The searches have been selected based on the number of leptons in the final state. Since the veto requirements are looser than the acceptance requirements for all the analyses performed, it is guaranteed that considering a 0-lepton, 1-lepton, 2-lepton SS and 2-lepton OS search, events are non-overlapping. This means that the events are statistically independent and hence multiplying the likelihood of all the searches together is justified.
 \item For backgrounds, the impact of the correlated systematics is small. Each considered search has a different set of backgrounds (by virtue of the chosen final state kinematics) and hence the background estimates are also uncorrelated between the different searches since they are probing different final states. There are some common elements, but these have been checked to be small. In any case, the background yields we consider (as provided by the searches) already have such systematics folded in and we also apply a 30\% uncertainty on all signal model points (see Sect.\ref{sec:validation}). 
\end{itemize}

\end{itemize}

Many other searches of vector-like quarks are present in literature, and they will be implemented in forthcoming upgrades of the code. For example, same-sign dilepton final states were accounted for in \cite{Chatrchyan:2013wfa}, searching for exotic partners with electric charge $5/3$ decaying exclusively to $W$ bosons and tops, excluding $m_{X}<770$ GeV\ at the 2$\sigma $ level. For the same integrated  luminosity at $\sqrt{s}=8$ TeV, the multilepton search \cite%
{CMS:2013una} excludes $B$ quark masses below 520-785\ GeV for non-nominal branching fractions to $Wt$, $Zb$ and $Hb$. The present ATLAS\ direct limits also reach mass bounds of 585 (645)\ GeV for the $T$ ($B$) vector-like
singlet scenario with 14\ fb$^{-1}$ of data, while a mass bound of 680 (725) GeV is now excluded at the 95\% CL for the branching ratio assumptions corresponding to weak-isospin doublets \cite{ATLAS:2013ima,TheATLAScollaboration:2013jha,TheATLAScollaboration:2013sha,TheATLAScollaboration:2013oha}. Yet, it must be kept in mind that such priors do not include the possibility for new partners having large couplings to the $u$, $d$, $c$ and $s$ quarks. While they fully cover the parameter space corresponding to top partners, non-zero mixings with the first two SM generations remain a likely possibility, and require careful treatment.

\subsection{Code restrictions}
\label{issues}

So far we have considered a simple implementation of XQ production and decays. 
However, there are several effects, not included in the present version of the code, that may affect the calculation of eCLs in our approach.
The main point in our framework is  to establish the exclusion in a
conservative and robust way. Therefore, it is extremely important to
identify and deal with all possible effects that can reduce the number of
predicted signal events. Conversely, an over-conservative estimate
would result in too weak bounds, so it is also relevant to
take into account any enhancing effect. The main factors which 
could affect the conservative estimate of the number of signal events and the 
respective limits are the following.

\begin{itemize}
\item \underbar{\textit{Interpolation of mass points}} Efficiencies have been computed only for a limited number of XQ masses (every 100 GeV). When computing the eCL for a XQ mass between two simulated values, we interpolate the result by relying on several methods, as described in detail in Appendix~\ref{app:eCL}. We have checked that they lead to similar results.

\item \underbar{\textit{Chain decays between XQs}} As discussed previously, decays like $Q \to Q' V$ (where $V$ is any SM boson, $W$, $Z$ or $H$) have not been included in the analysis. In principle, their inclusion is straightforward, even though it would require a scan over two masses. However, we have decided not to include them in order to keep the tool simple and also because such decays are only relevant when there are large mass splittings between the two XQs. Furthermore, even when kinematically allowed, decays directly to SM states tend to always dominate when a sizeable  mixing to the SM quarks is allowed, as it is common in explicit models.

\item \underbar{\textit{Decays into other states in the model}} Further, decays like $Q \to q V_{BSM}$, where $V_{\rm BSM}$ is a new boson present in the model the XQ belongs to, have not been included either as they are model dependent and it would not be easy to implement a complete set of $V_{\rm BSM}$ candidates. Also, typical mass limits on  $V_{\rm BSM}$ states may be higher than those on XQs (especially if their leptonic decays are not suppressed, in which case they can be accessed in Drell-Yan processes), so that such decays are not kinematically possible.

\item \underbar{\textit{Interference effects}} In the presence of multiple XQs, there is the possibility that the decays of different states would lead to identical final states, therefore interference between the various signal processes may occur. This effect is not included in our analysis, where we need to separate  the various channels.
A first quantitative estimate of these effects is provided in\cite{Barducci:2013zaa} and a full treatment will be included in next upgrades of the code.

\item \underbar{\textit{Loop corrections to masses and mixing}} A potentially relevant effect comes from EW corrections to the masses and mixings of the XQs, as they may, for instance, remove or add degeneracies, or change the BRs. However, such effects are highly model dependent and it is left to the user to check whether they are relevant in the model of interest:  loop corrected masses and BRs can be provided as input to \verb|XQCAT|. A detailed and quantitative treatment of this dynamics will be done by applying the technique proposed in \cite{Cacciapaglia:2009ic} to fermion propagators and its implementation will be considered in a future upgrade of the code.

\item \underbar{\textit{Higher order cross section}} The pair production cross section receives sizeable QCD corrections. Under the approximation that the kinematics is unaffected by the latter, the effect can be added via a model independent $k$-factor. Therefore, we have considered for our simulation the cross sections computed at Next-to-Leading Order supplemented by Next-to-Next-Leading-Logarithmic resummation
(NLO-NNLL) in QCD in \cite{oai:arXiv.org:1111.5869}. EW loop corrections may also be relevant, however, they are model dependent and they are expected to be smaller than the QCD ones. For this reason, we do not consider them here.

\end{itemize}

\section{Validation of the framework}
\label{sec:validation}

The validation of our tool for XQ searches has been done by comparing our results to experimental data for some specific channels. For this purpose, we have considered the CMS inclusive search \cite{Chatrchyan:2013uxa,twikiB2G12015} and analysed the same benchmark points (i.e., $T$ masses and BRs) considered in the search. 
The validation has the purpose of testing the two main sections of our framework: the limit code that computes the eCLs starting from the number of signal events obtained from input and the code that extracts the efficiencies considering the selection and kinematics cuts of the implemented searches. We dwell on this below while we refer the reader to \cite{Buchmueller:2013exa} for the
case of SUSY searches.

\subsection{Validation of the limit code} 

The limit code has been tested by computing the expected and observed limits using the information provided in the 
experimental search documentation. This test allows us to determine any discrepancies between the statistical method used in our approach and the one in the CMS analysis. The resulting eCLs and 95\% mass bounds computed with \verb|XQCAT| considering the combination of all channels or of the multi-lepton channels only are shown in Tab.\ref{tab:validationlimitcode}. The uncertainty on the signal events has been assumed to be 20\%. The results of this test show two different effects: 

\begin{enumerate}
\item due to a different analysis technique, we are not able to reproduce the mass bounds considering the single lepton-channels in combination with the multi-lepton channels; 
\item considering only the multi-lepton channels we can reproduce the experimental expected (observed) mass bounds with a discrepancy of $-8\%$ ($-6\%$).
\end{enumerate}

For these reasons, we will only consider the multi-lepton channels in the implementation of this XQ search in our framework. Similar considerations can be done for all the implemented searches.

\begin{table}
\scriptsize
\centering
\setlength{\tabcolsep}{2pt}
\begin{tabular}{|c|cc|cccc|cc|cc|}
\cmidrule{1-7}                                
& \multicolumn{2}{c|}{Single-lepton channels} & \multicolumn{4}{c|}{Multi-lepton channels} \\
\cmidrule{1-7}                                
& Muon & Electron & OS1 & OS2 & SS & 3l \\
\cmidrule{1-7}                                
Bg & $61900\pm13900$ & $61500\pm13700$ & $17.4\pm3.7$ & $84\pm12$ & $16.5\pm4.8$ & $3.7\pm1.3$ \\
Data & 58478 & 57743 & 20 & 86 & 18 & 2 \\
\midrule
\multicolumn{7}{|c|}{\multirow{2}{*}{Signal events for nominal point (BR$(Wb)=0.5$ and BR$(Zt)={\rm BR}(Ht)=0.25$)}} & \multicolumn{2}{c|}{${\rm eCL}_{\rm\tiny all}$} & \multicolumn{2}{c|}{${\rm eCL}_{\tiny\mbox{multilepton}}$} \\
\multicolumn{7}{|c|}{} & Exp & Obs & Exp & Obs \\
\midrule
500 GeV  & 850  & 840  & 16.7 & 35.1 & 21.3 & 19.1 & 0.29 & 0.36 & 1     & 1 \\
600 GeV  & 280  & 280  & 8.9  & 16.6 & 7.5  & 8.5  & 0.20 & 0.25 & 0.998 & 0.999 \\
700 GeV  & 97   & 98   & 4.0  & 6.6  & 2.8  & 3.1  & 0.10 & 0.13 & 0.831 & 0.851 \\
800 GeV  & 36   & 37   & 1.6  & 2.5  & 1.0  & 1.3  & 0.05 & 0.06 & 0.438 & 0.487 \\
\midrule                                
\midrule                                
\multicolumn{7}{|c|}{95\% exclusion limit computed by the limit code} & - & - & 626 GeV & 630 GeV \\
\bottomrule
\end{tabular}
\caption{eCLs  and 95\% lower mass bounds obtained with the statistical combination of search bins implemented in the limit code. The quoted values for the expected (observed) lower mass bounds in \cite{Chatrchyan:2013uxa} are 773 GeV (696 GeV) considering all channels, and 683 GeV (668 GeV) considering only the multi-lepton channels (as reported in the corresponding twiki page \cite{twikiB2G12015}, and to be compared to the limit code results in the table).}
\label{tab:validationlimitcode}
\end{table}

\subsection{Validation of the efficiency extraction code}

The extraction of the efficiencies depends on the interplay of different parameters: the most relevant ones are the accuracy of the MC simulation, the correct reproduction of the true detector effects using a fast detector simulation and the correct reproduction of the experimental selection and kinematic cuts. The number of events in the multi-lepton channels for different values of the $T$ mass computed in our framework is shown in Tab.\ref{tab:efficiencytable}. Comparing these values to the numbers in Tab.\ref{tab:validationlimitcode} it is possible to notice that our discrepancies are almost always within $\pm$10\% in all the multi-leptonic channels, except for the second opposite-sign di-lepton channel (OS2), where the discrepancy is consistently larger. The differences between our computed number of events and the values quoted in the experimental search can be explained by unavoidable differences in the modeling of the detector and in the implementation of the selection cuts. A further exploration of 
these discrepancies would require a more precise knowledge of the details of the measurements and a more accurate simulation of detector effects, which are not possible with the information and tools currently available. For this reason, we have decided to omit the OS2 channel from the implementation as -- in essence -- it cannot be accurately reproduced. The channels used to extract our results with the search \cite{Chatrchyan:2013uxa} are only the first opposite-sign di-lepton (OS1), the same-sign di-lepton (SS) and the tri-lepton (3l) channels. The uncertainty in the number of events has been set to 20\% to take into account both the uncertainty in the efficiencies and the uncertainty in the NLO-NNLL production cross section computed following \cite{oai:arXiv.org:1111.5869}.

\begin{table}
\small
\centering
\setlength{\tabcolsep}{2pt}
\begin{tabular}{|c|cccc|}
\toprule
Mass & OS1 & OS2 & SS & 3l \\
\midrule
500 GeV  & 16.9 (+1.2\%) & 20.5 (-42\%) & 20.2 (-5.2\%) & 20.5 (+7.3\%) \\
600 GeV  & 9.6  (+7.9\%) & 10.3 (-38\%) & 8.2  (+9.3\%) & 9.1  (+7.1\%) \\
700 GeV  & 4.0  (0\%)    & 4.4  (-33\%) & 2.8  (0\%)    & 3.7  (+19\%) \\
800 GeV  & 1.7  (+6.3\%) & 1.5  (-40\%) & 0.8  (-20\%)  & 1.2  (-7.7\%) \\
\bottomrule                                
\end{tabular}
\caption{Number of signal events simulated in our framework for the nominal point with BR$(Wb)=50\%$ and BR$(Zt)={\rm BR}(Ht)=25\%$. In parenthesis, the relative discrepancy with the number of events quoted in \cite{Chatrchyan:2013uxa}.}
\label{tab:efficiencytable}
\end{table}

\subsection{Analysis of one $T$ singlet mixing only with Standard Model top}

As a final step to the validation of our tool and of the implementation of an experimental search, we used \verb|XQCAT| to compute the 95\% CL mass bound for a $T$ singlet under different hypotheses for its BRs into third generation quarks and SM bosons. All the results presented in this {and the following} sections assume as an input the
NLO-NNLL order cross sections for $pp\rightarrow Q\bar{Q}$ production at the LHC, as given in \cite{oai:arXiv.org:1111.5869}. 

In Fig. \ref{fig:validationTsingletthirdgen} we show the eCLs for a $T$ singlet with BR$(Wb)=50\%$ and BR$(Zt)={\rm BR}(Ht)=25\%$. To extract a mass bound we use two of the methods described in the Appendix~\ref{app:eCL}: with a linear interpolation of the eCLs we obtain a 2$\sigma$ mass bound of 643 GeV while a linear interpolation of the efficiencies yields a 2$\sigma$ mass bound of 674 GeV. These results show that the numerical value of the mass bound is not very sensitive to the interpolation method in use, at least in a typical situation. 
Our numerical results are summarised in Tab.~\ref{tab:numericalbounds} of the Appendix~\ref{app}. These values can be compared with the quoted value of 668 GeV in the multi-lepton channels only \cite{twikiB2G12015}. In Fig.\ref{fig:validationTsingletthirdgen} we have also plotted the eCLs obtained from the combination of all SUSY searches, at $\sqrt{s}=7$ and 8 TeV. This comparison will be performed in all cases when the difference in sensitivity of the direct search and of the searches inspired by SUSY scenarios for specific final states can play a crucial role 
in determining the eCLs. In such a case, we observe that the direct search is more sensitive than the SUSY combination. The bound provided by linearly interpolating the {\rm eCL}s of the SUSY searches combination is 563 GeV (590 GeV if interpolating the efficiencies). Taking into consideration that SUSY searches were not designed to be sensitive to this kind of final states, it is remarkable that the obtained bound is not too far from the one reproduced by the direct XQ search. We therefore highlight the important role that the SUSY searches may have so as to explore scenarios where {XQ}s do not decay to heavy generations and for which direct searches of {XQ}s (that usually require $ b$-jets in the final states) might not be as sensitive. We will further explore this possibility in the following section.

To further validate our implementation, we computed the $T$ mass bounds varying all the BRs between 0 and 1 in steps of 0.2.
Our results in the BR$(Wb)$-BR$(Ht)$ plane are shown in Fig.~\ref{fig:validationtriangle}. Again, to be able to compare with the experimental results, the 95\% CL contours on the $T$ quark mass are obtained by linear interpolation of the eCLs between the simulated points. The comparison with the experimental values of the CMS search in the multi-lepton channels \cite{twikiB2G12015} shows that our results are consistent with the experimental values within less than 40 GeV (corrisponding to less than $\pm$10\%) for most BRs configurations.

\begin{figure}
\centering
\epsfig{file=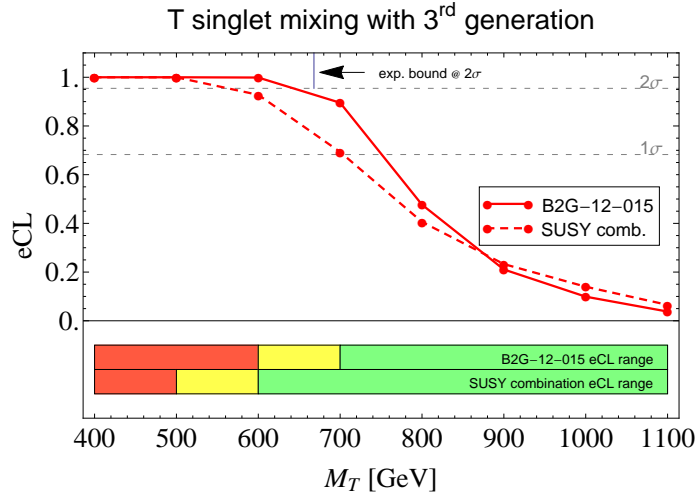,width=.6\textwidth}
\caption{Exclusion confidence levels for a $T$ singlet mixing only with the top quark, that is with BR$(Zt)={\rm BR}(Ht)=0.25$ and BR$(Wb)=0.5$. The dots correspond to the simulated points, while the lines are linear interpolations of the {\rm eCL}s (method 3 in App.\ref{app:eCL}). The solid line corresponds to the {\rm eCL}s obtained using the direct search \cite{Chatrchyan:2013uxa}, while the dashed line corresponds to the combination of the SUSY searches at $\sqrt{s}=7$ and 8 TeV. The strips below the plot correspond to method 2 of App.\ref{app:eCL}. The red region is excluded at 95\% CL, the yellow region is where the $2\sigma$ eCL can be found, the green region is not excluded at 95\% CL.}
\label{fig:validationTsingletthirdgen}
\end{figure}

\begin{figure*}
\centering
\subfigure[]{\epsfig{file=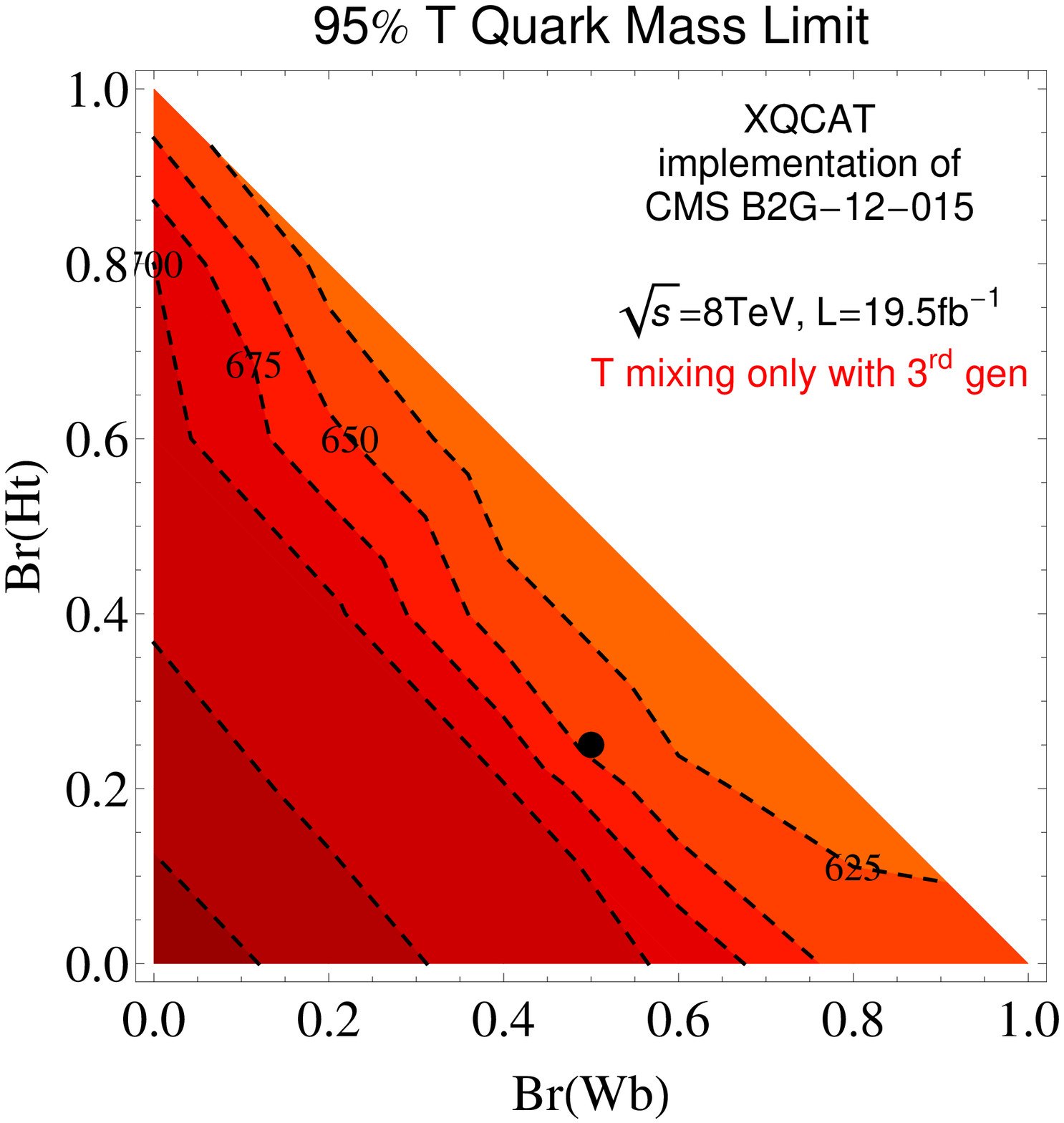,width=.48\textwidth}}\hfill
\subfigure[]{\epsfig{file=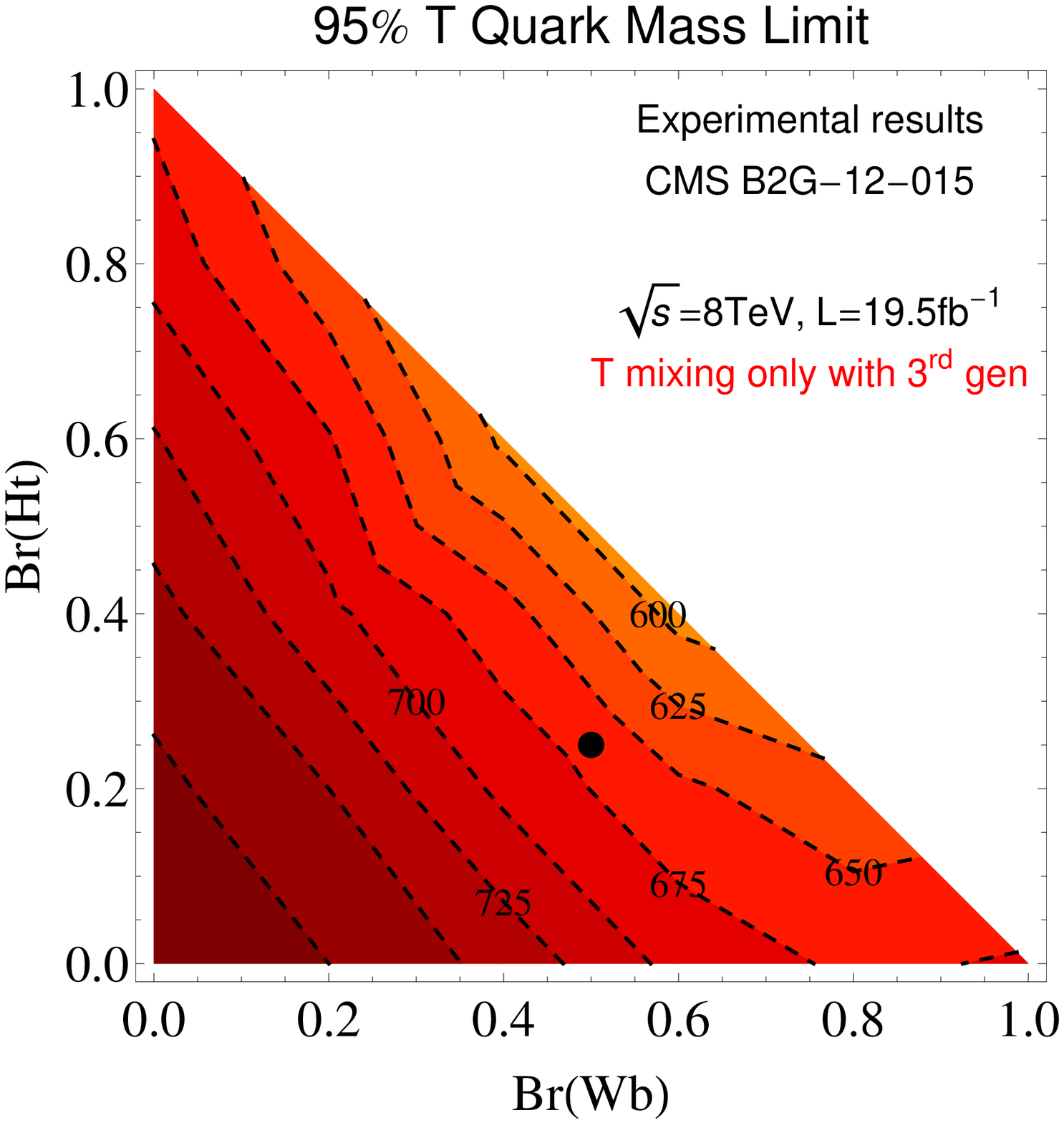,width=.48\textwidth}}
\subfigure[]{\epsfig{file=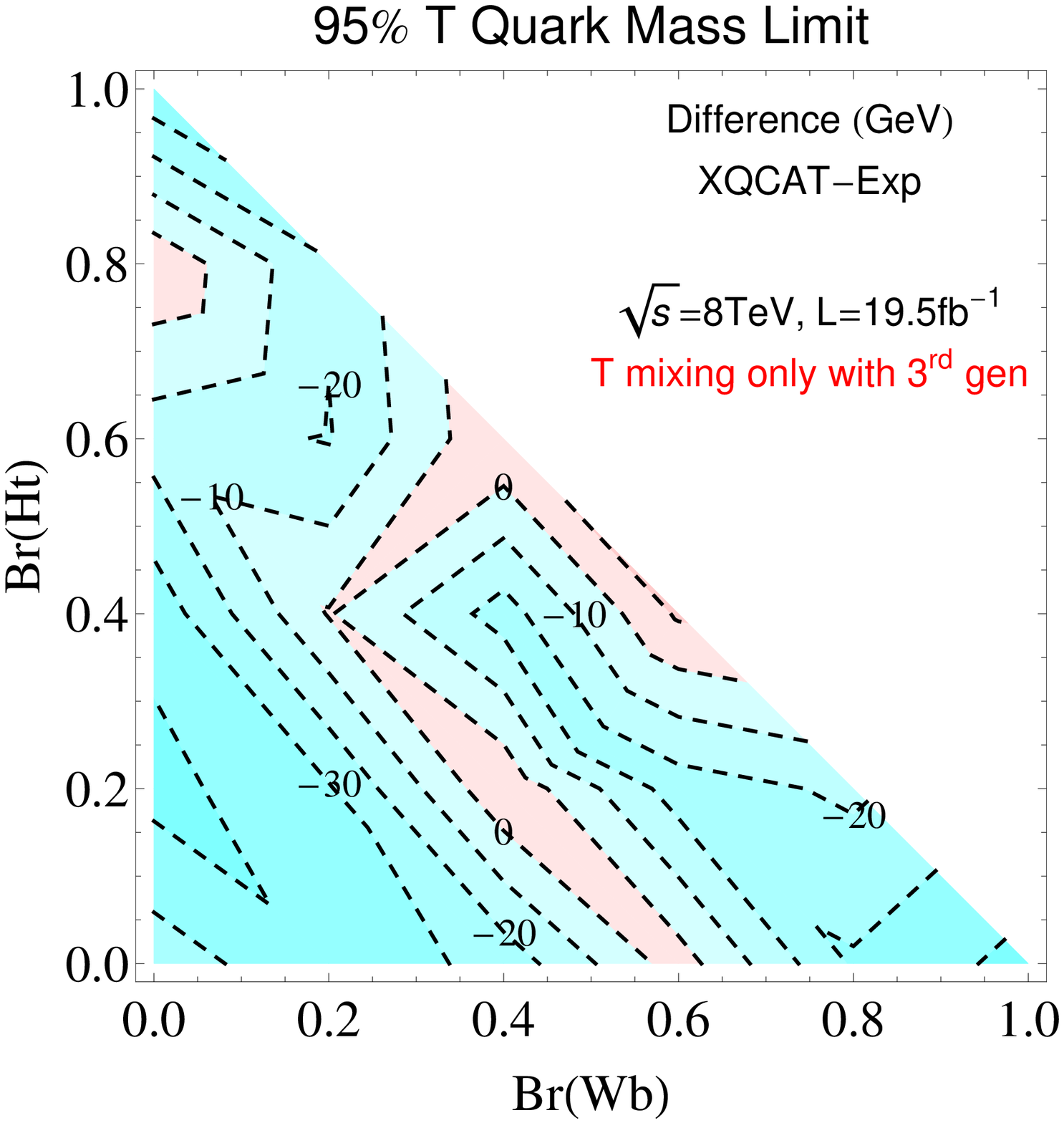,width=.48\textwidth}}\hfill
\subfigure[]{\epsfig{file=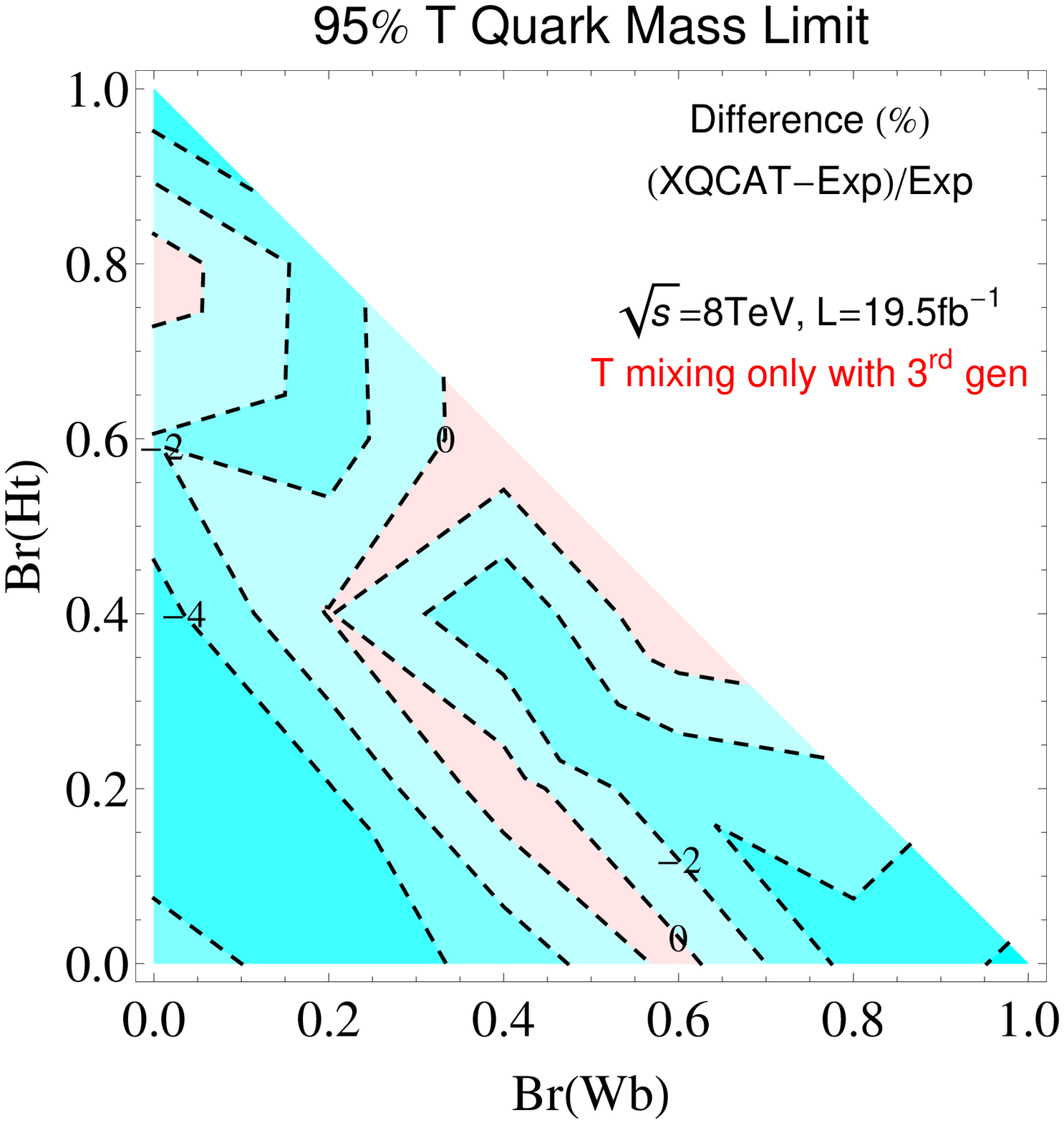,width=.48\textwidth}}
\caption{95\% CL mass limit on the $T$ quark mass in the BR$(T\to Wb)$-BR$(T\to Ht)$ plane for the results obtained with {\tt XQCAT} (a) and the experimental results of \cite{twikiB2G12015} (b). The 95\% contours have been obtained through linear interpolation of the eCLs for the simulated masses, i.e. from 400 GeV to 1000 GeV with steps of 100 GeV. The black dot represents the nominal point for a $T$ singlet with ${\rm BR}(Zt)={\rm BR}(Ht)=0.25$ and BR$(Wb)=0.5$. The differences between our results and the experimental results are shown in (c) (difference in GeV) and (d) (relative difference in \%). }
\label{fig:validationtriangle}
\end{figure*}

\section{Constraints on scenarios with multiple XQs}
\label{sec:MultiXQ}

In this section, we test the ability of \verb|XQCAT| to set bounds on scenarios with many XQs in toy models as well as more realistic scenarios. Re-interpreting the experimental results on XQs in models where more than one new fermion
is present is not an easy task, as already intimated. It is even more involved when the XQs are degenerate in mass. 
The main difficulty in such re-interpretation of the (single XQ) searches is found in the fact that signals coming from different states can contribute to the same signal bin, therefore one needs to calculate the efficiencies and number of events that pass all the cuts before calculating the eCL.
This is exactly what our program \verb|XQCAT| does.

In general, signal bins can receive contributions from different physical states in the following two cases:
\begin{itemize}
\item[-] the model contains two states with the same decay channels which therefore produce the same final state;
\item[-] the model contains states with different decay channels, however, the signal features are (even  partially) sensitive to the different final states.
\end{itemize}
To illustrate how the combined limits affect the excluded parameter space, we will here consider XQs which
 mix only to the third generation and exploit the bounds given by the CMS B2G-12-015 searches \cite{Chatrchyan:2013uxa}, which are specifically designed for pair produced $T$ vector-like XQs decaying to $Wb$, $Zt$ and $Ht$.

For the purpose of setting representative bounds on models with extended quark sectors, one can consider two
simplified scenarios:
\begin{itemize}
\item two $T$ singlets;
\item a doublet $\{X,T\}$ with exotic hypercharge, $Y=7/6$.
\end{itemize}
The two cases above can be described in terms of 3 free parameters: the mass of the XQ multiplet, the coupling strength and a parameter encoding the BR into the third generation quarks.
The numerical implementation of the models used here can be found in FeynRules \cite{FeynRules,FeynRulesVLQ}
and HEPMDB~\cite{hepmdb, hepmdbVLQ}.

 The $T$ singlet and the $\{X,T\}$ doublet scenarios above were identified in \cite{Buchkremer:2013bha} as being among the
least constrained XQ representations from flavour observables, whenever assuming general
mixing with the three SM quark families.

Let us start with the first simplified scenario of two $T$ singlets with arbitrary mass. Both states have similar decays: $\sim 50\%$ in $Wb$ and $\sim 25\%$ in $Zt$ and $Ht$. 
Using these  BRs as inputs to the code, we calculated the eCLs for each value of the two masses $\{m_{T_1}, m_{T_2}\}$.
The result is shown in Fig.\ref{fig:2Tsinglets}.

\begin{figure}
\centering
\subfigure[]{\epsfig{file=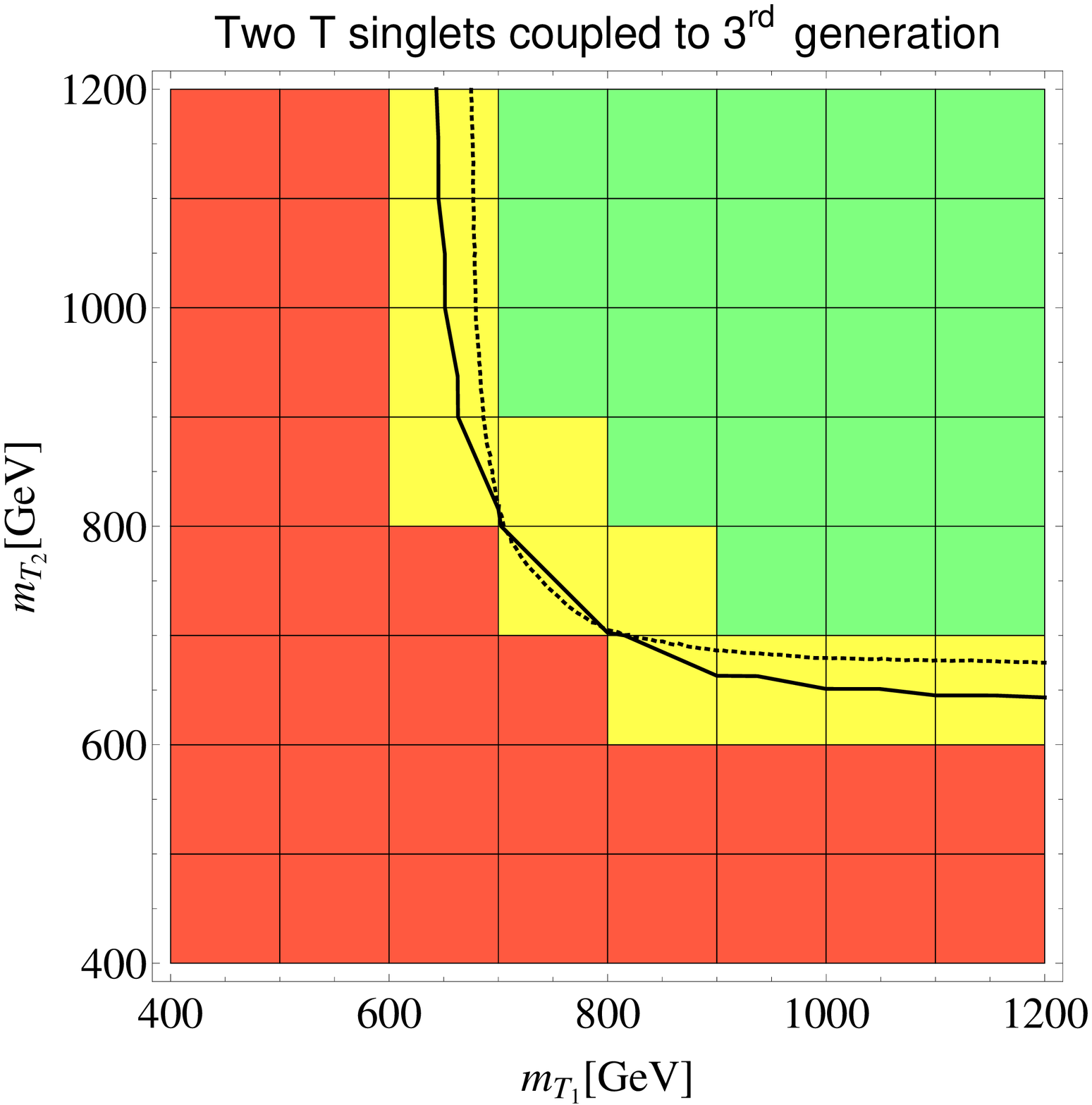,width=.48\textwidth}\label{fig:2Tsinglets}}
\subfigure[]{\epsfig{file=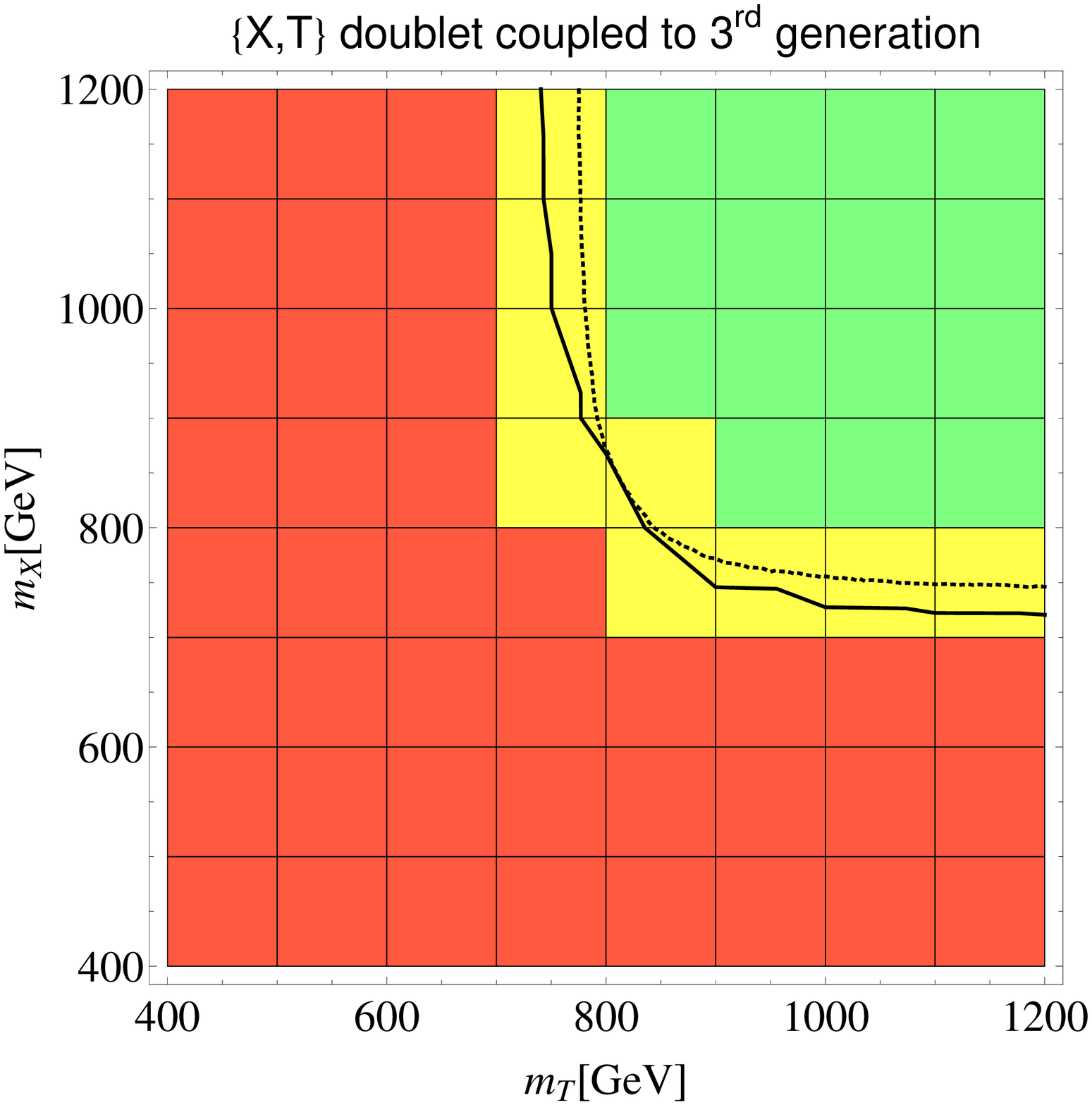,width=.48\textwidth}\label{fig:XTdoublet}}
\caption{\label{fig:multiXQthird}Here we show eCLs considering the CMS search B2G-12-015 \cite{Chatrchyan:2013uxa} for a configuration of (a) 2 $T$ singlets or (b) $\{X, T\}$ doublet with masses ranging between 400 GeV and 1200 GeV coupled only to third generation SM quarks. The corresponding BRs are given by Eqs.\ref{Eq:AtreBR} and \ref{Eq:AlvesBR}, respectively. The excluded (red), boundary (yellow) and non-excluded (green) regions at 95\% CL are shown. The solid (dashed) contour corresponds to the 95\% CL bound obtained by linear interpolation of the eCLs (efficiencies) for the simulated points.}
\end{figure}

The crossing points in the grid of the figure correspond to the mass values that we simulated.
The most conservative way to set bounds is to calculate the eCLs on the simulated masses, which are therefore fully reliable.
Shown in red on the plot are the squares whose corners are all excluded at 95\% CL, while in green the squares whose corners are allowed.
The yellow regions contain the intermediate situation, i.e.,  squares where only some of the corners are excluded: we can then affirm that the exclusion limit must be a line crossing the yellow squares.
This is proven by the two black lines: the solid one corresponds to the bound one obtains by interpolating the eCL calculated on the simulated points, as it is usually done in experimental results. 
As a check, we used the code to do a finer scan on the masses, using efficiencies which are interpolated between simulated points. This procedure, in general, gives rise to different values for the bounds, nevertheless we see that the corresponding dashed line falls very close to the solid one.
This plot displays interesting physics results: it shows that the obtained bound is mostly sensitive to the presence of the two states when their mass differs by less than 200~GeV\footnote{As stated, we are not including interference effects in this analysis. They could be relevant when the XQ's are degenerate enough i.e. for values close to the diagonal in the plots in Fig.\ref{fig:multiXQthird}, and also in Figs.\ref{fig:Rattazzi},\ref{fig:multiXQfirst}.}. In contrast, for larger mass differences, the presence of the second heavier state becomes irrelevant and the bound coincides with the one obtained with one state only, i.e., in the 600--700 GeV range at 95\% CL.

To study the impact of the presence of XQs with different charge, we also present the case of a XQ doublet with 
non-standard hypercharge, $\{X,T\}$, which contains a charge $5/3$ state $X$ and a charge $2/3$ state $T$ ~\cite{Cacciapaglia:2011fx}.
The exotic charge state decays $100\%$ into $Wt$: this mode has been searched for in final states with two same sign leptons plus jets~\cite{ATLAS:2012hpa,Chatrchyan:2013wfa}, giving a bound of 670 GeV (ATLAS, pair production only) or 800 GeV (CMS). Although it does not correspond to physically realistic situations as the mass splitting can only be generated by mixing via the Higgs boson, we consider the bounds for two masses ranging independently between $400$ to 1200~GeV. This scan is a good exercise to illustrate our point.
The result is shown in Fig.\ref{fig:XTdoublet}: the plot shows again that an interplay between the two masses emerges when the mass splitting is smaller than 200~GeV, like in the case of the two $T$ singlets.
However, the most striking result is that the searches for $T$ states can already pose bounds on the $X$ state.
Further, even in models where a single $T$ state is present, one cannot simply utilise directly the experimental bounds: the other partners in the same multiplet as $T$ can contribute to the signal rate and thus increase the bound on its mass. 

We conclude this section by studying two motivated models which contain multiple XQs and show the bounds on their parameter spaces as given by  our code.
The first model we consider describes the top partners which are usually added in models of pseudo-Goldstone (composite) Higgs, based on the minimal coset $SO(5)/SO(4)$.
The top partners are assumed to belong to a bi-doublet and singlet of $SO(4)$ \cite{DeSimone:2012fs}: here we focus on the bi-doublet, which is typically expected to be lighter than the singlet.
The $SO(4)$ fourplet is decomposed into the two $SU(2)_{L}\times U(1)_{Y}$ doublets \{T,B\} and $\{X_{5/3},X_{2/3}$\}. (Notice that we are here using the notation introduced by the authors of \cite{DeSimone:2012fs} for the  doublet with non-standard hypercharge $\{X,T\}$.)  
All the states are assumed to decay exclusively to third generation quarks, such that 
\begin{eqnarray}
&{\rm BR}(X_{5/3} \rightarrow W^{+}t)={\rm BR}(B\rightarrow W^{-}t)=100\%\nonumber&\\
&\left\{ 
\begin{array}{l}
{\rm BR}(X_{2/3}\rightarrow Zt)={\rm BR}(X_{2/3}\rightarrow Ht)=50\% \\ 
{\rm BR}(T\rightarrow Zt)={\rm BR}(T\rightarrow Ht)=50\%%
\end{array}%
\right.\nonumber&
\label{Eq:AtreBR}
\end{eqnarray}
In our study we assume degenerate masses for the components of the two doublets:  $m_{X_{5/3}}=m_{X_{2/3}}$ and $m_{T}=m_{B}$.
This is a natural approximation, since, as mentioned before, the mass splitting can only be generated by the Higgs vacuum expectation value and is therefore relatively small.
However, the mass difference between the two doublets can be large.
In Ref.\cite{DeSimone:2012fs}, it is argued that the $\{X_{5/3}, X_{2/3}\}$ doublet is lighter and that the mass splitting with the other doublet is rather large, being proportional to the composite scale times the top Yukawa coupling. 
In general, however, the mass difference can be small depending on the specific model. For instance, in models based on a warped extra dimension~\cite{Contino:2006nn,Hosotani:2008tx}, the difference in mass between the two doublets is controlled by the different boundary condition on the Planck (ultra-violet) brane. It is well known that this splitting varies depending on the localisation of the zero modes, which are identified with the top and bottom left-handed doublet.
If the left-handed tops are localised towards the TeV brane, thus being ``composite'' in the AdS/CFT interpretation, the mass splitting is indeed large and of the same order as the mass. Conversely, if the left-handed top is close to flat or localised on the Planck brane (i.e., mostly elementary), the mass difference drops to zero.
In our study, we let the two masses vary independently between $400$ to $1200$ GeV.
The results are shown in Fig.\ref{fig:Rattazzi3}.  As expected, the bounds are much more stringent in the region of small splitting between the two doublets and can be set in the 800--900 GeV range.

Our code \verb|XQCAT| is not limited to vector-like fermions and can also be applied to models with chiral exotic fermions, like fourth generation ones.
To illustrate this application, we focus on the model proposed  in Ref.\cite{Alves:2013dga}.
It consists of two Higgs doublets with a new set of anomaly free chiral fermions.
The new exotic states involve two separate left-handed doublets $\{X,T\}$ and 
\{B,Y\}. (Again we use the notation introduced by the authors of \cite{Alves:2013dga} for the  new doublets). Imposing an appropriate discrete symmetry,  all the Flavour Changing
Neutral Current (FCNC) interactions are suppressed at the LHC energies, whereas
the new resonances are allowed to mix with the third generation quarks with
the rates 
\begin{eqnarray}
&&{\rm BR}(X \rightarrow W^{+}t)={\rm BR}(T\rightarrow W^{+}b)=100\%,  \notag \\
&&{\rm BR}(B \rightarrow W^{-}t)={\rm BR}(Y\rightarrow W^{-}b)=100\%. \nonumber
\label{Eq:AlvesBR}
\end{eqnarray}
The results shown in Fig.\ref{fig:Alves3} indicate that, also in the case of exotic fermions, if their masses are of the same order, the exclusion bounds are larger.

\begin{figure}
\centering
\subfigure[]{\epsfig{file=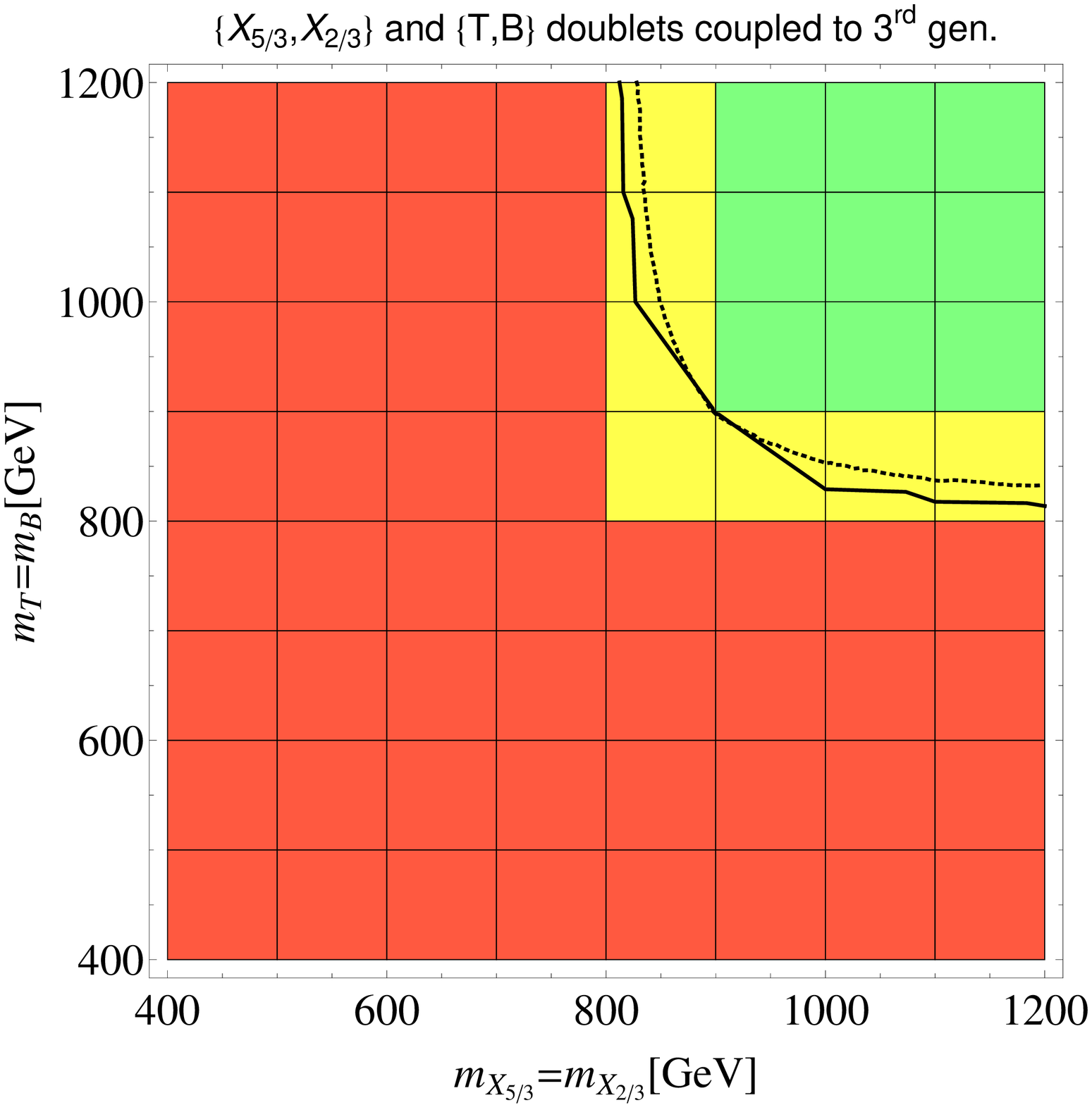,width=.48\textwidth}\label{fig:Rattazzi3}}
\subfigure[]{\epsfig{file=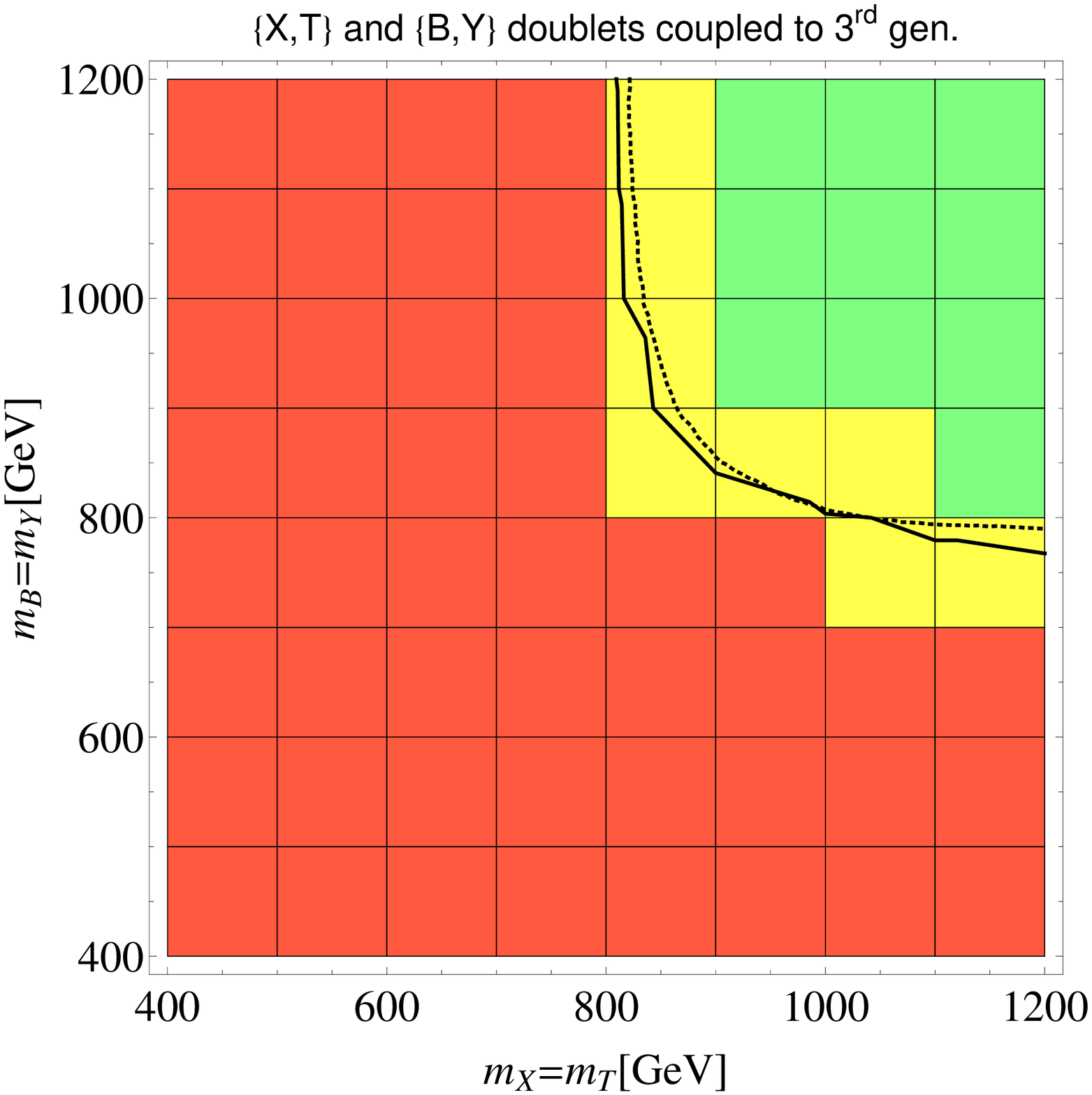,width=.48\textwidth}\label{fig:Alves3}}
\caption{95\% CL excluded (red), boundary (yellow) and non-excluded (green) regions for simplified versions of (a) the $\{X_{5/3},X_{2/3}\}$ and $\{T,B\}$ doublets in  model \cite{DeSimone:2012fs} plus (b) the  $\{X,T\}$ and \{B,Y\} doublets in  the  model of  \cite{Alves:2013dga}, both with exclusive couplings to the third generation. The solid (dashed) contour corresponds to the 95\% CL bound obtained by linear interpolation of the eCLs (efficiencies) for the simulated points.}
\label{fig:Rattazzi}
\end{figure}

\section{Interplay and complementarity with other searches}

Another important issue that we want to address with the \verb|XQCAT| code is the complementarity between direct searches for XQs and other searches performed at the LHC.
As a first exploration we have focused on SUSY searches, which are fully implemented in the code.
From our validation, we have already shown in Sect.~\ref{sec:validation} (see for instance Fig.~\ref{fig:validationTsingletthirdgen}) that SUSY searches can give bounds on XQ masses which are close to the direct searches in the case of exclusive mixing to the third generation.
In this Section we will focus on the scenario where the XQs mix with the light generation quarks: this case has recently received great attention by the experiments, as it may give rise to large single production cross sections~\cite{Atre:2011ae}.
It has also been shown that the flavour bounds do not disfavour cases where significant mixing to both the top and either the up or the charm quarks is turned on~\cite{Cacciapaglia:2011fx}.
Nevertheless, no specific search focused on pair production followed by decays to light jets is available and here we will show that SUSY data samples can already set significant bounds.

\begin{figure}
\centering
\subfigure[]{\epsfig{file=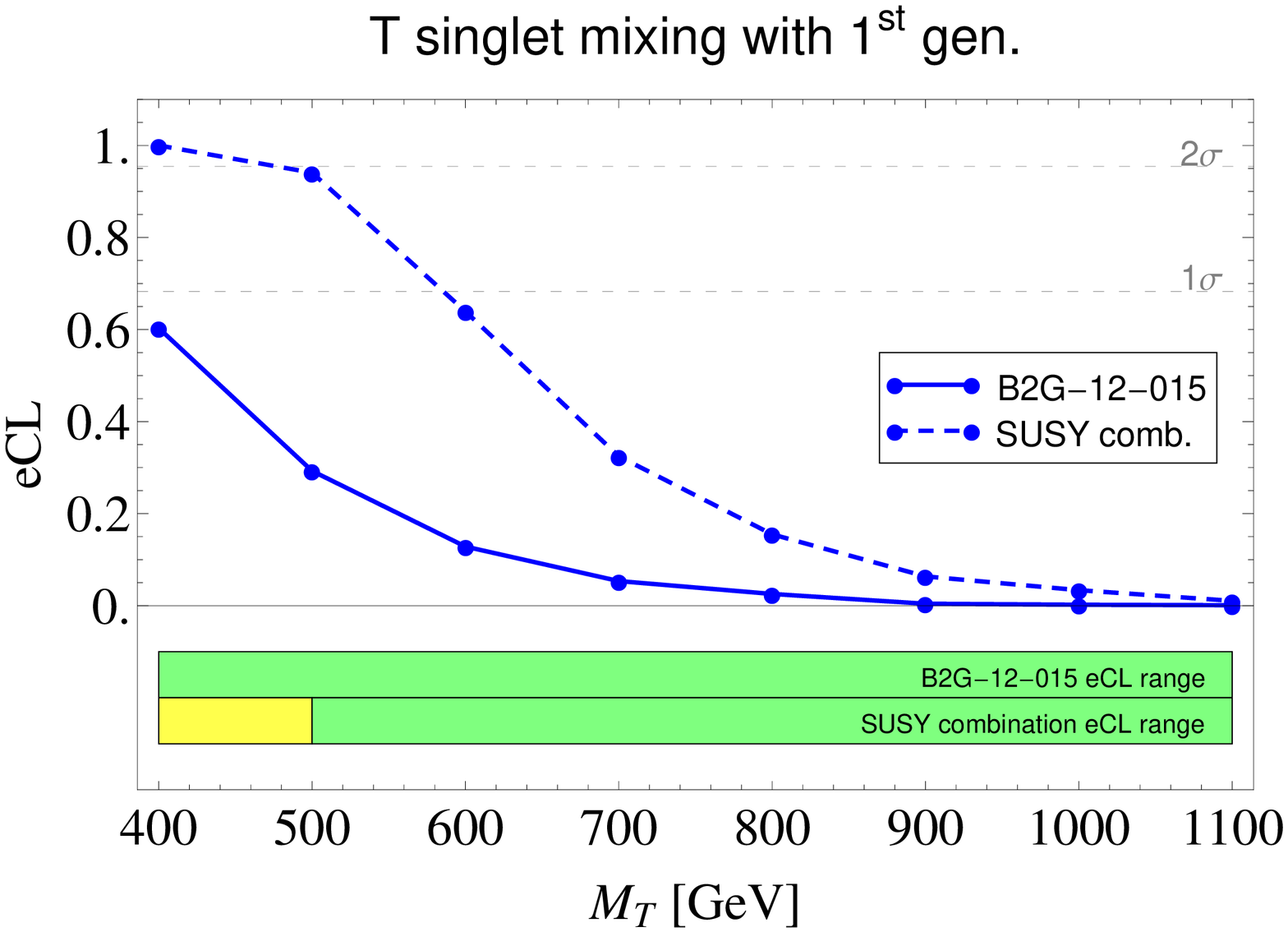,width=.48\textwidth} \label{fig:validationTsingletfirstgen}}
\subfigure[]{\epsfig{file=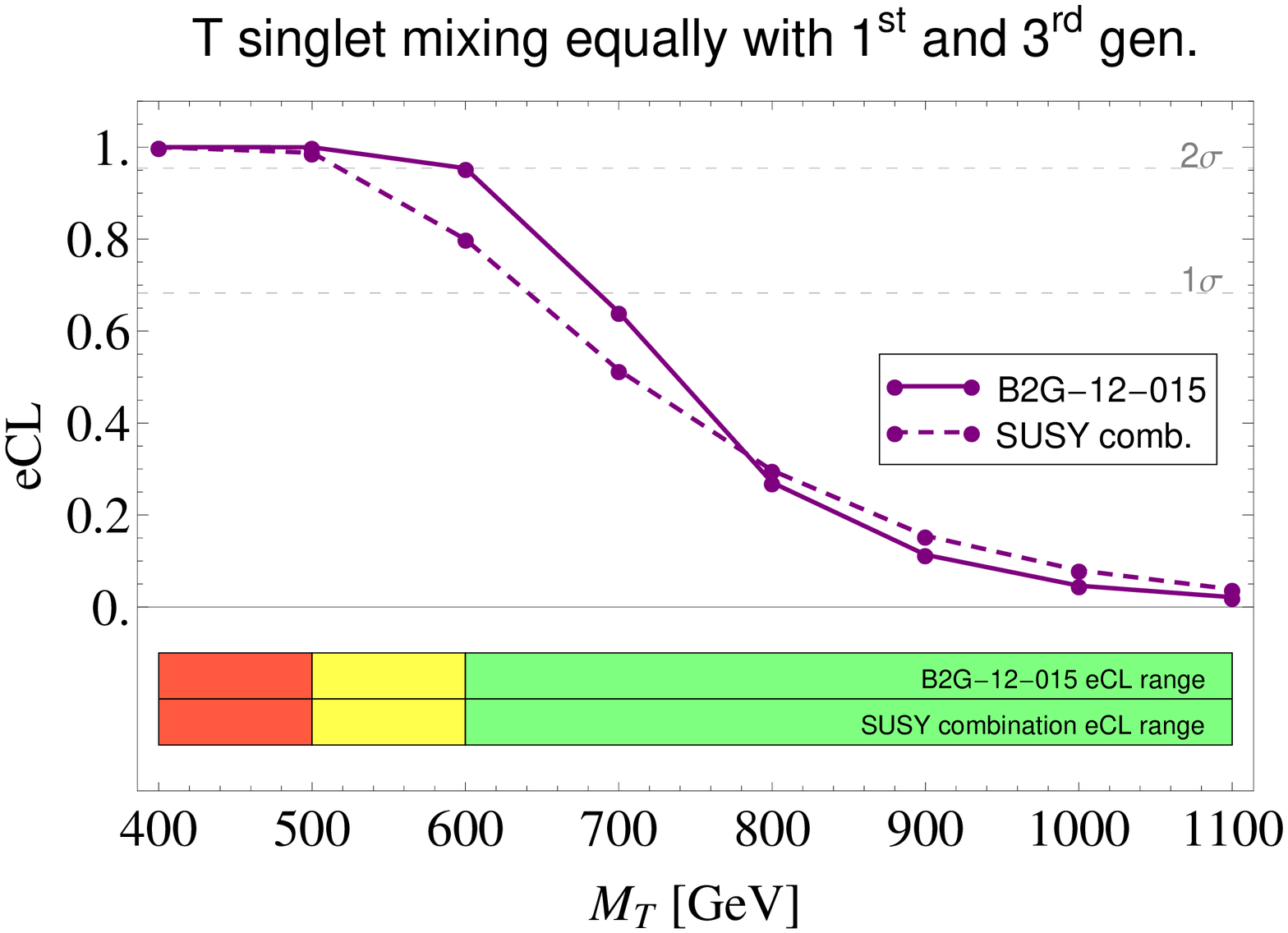,width=.48\textwidth}\label{fig:validationTsingletfirstthirdgen}}
\caption{Here, we show eCLs for (a) for a $T$ quark coupled only to the SM up quark such that ${\rm BR}(Zu)={\rm BR}(Hu)=0.25$ and BR$(Wu)=0.5$ and (b) for a $T$ quark mixing equally with the first and the third generations, with ${\rm BR}(Wb)={\rm BR}(Wd)=0.25$ and ${\rm BR}(Zt)={\rm BR}(Zu)={\rm BR}(Ht)={\rm BR}(Hu)=0.125$. The dots correspond to the simulated points while the lines are linear interpolations of the {\rm eCL}s (method 3 in App.~\ref{app:eCL}). The strips below the plot correspond to method 2 of App.~\ref{app:eCL}. The red region is excluded, the yellow region is where the 95\% eCL can be found and the green region is not excluded.}
\end{figure}

In Fig. \ref{fig:validationTsingletfirstgen}, we first consider the simple case of a singlet $T$, which is allowed to mix with light quarks only, with the following 
decay rates: ${\rm BR}(Wj)=0.5$ and ${\rm BR}(Zj)={\rm BR}(Hj)=0.25$.
Following the same method as in Sect.~\ref{sec:validation}, we find, as expected, that the sensitivity of the B2G-12-015 direct searches are strongly reduced: this is not a surprise as the final state loses many jets and leptons from the top-decays and is depleted of $b$-jets, which are used systematically to tag the signal region.
It is remarkable that, despite a sensible drop in
sensitivity, the combination of the SUSY searches still yields a 95\%
eCL mass limits above $400$ GeV, compared to the case of dominant
coupling to third generation quarks. In the Appendix~\ref{app}, we display in Tab.~\ref{tab:numericalbounds} the corresponding mass limits for different combinations of the BRs. The obtained results indicate that the SUSY searches combined at $\sqrt{s}=7$ and 8 TeV lead to a mass bound in the ballpark $400$--$500$ GeV, except in the case of exclusive decays into Higgs boson. Fig.~\ref{fig:validationTsingletfirstthirdgen} displays the sensitivity of the searches if mixed decay modes are present, e.g., if equal couplings to the up the top quarks are allowed. In this specific situation, corresponding to the configuration ${\rm BR}(Wb)={\rm BR}(Wd)=0.25$ and ${\rm BR}(Zt)={\rm BR}(Zu)={\rm BR}(Ht)={\rm BR}(Hu)=0.125$, the direct search \cite{Chatrchyan:2013uxa} displays a much better sensitivity compared to the previous case, while the combination of the SUSY searches sets a slightly milder constraint, though still in the same mass range (500-600 GeV) of the direct search.

From a channel-by-channel analysis, the interpolated {\rm eCL}s and efficiencies indicate that the most sensitive handle on $T$ partners coupling exclusively to the light generations consists in looking for XQs decaying to $Z$ bosons and jets. This implies that the forthcoming
searches for pair-produced XQs would certainly benefit from the inclusion of
such decay modes, as the assumption BR$(Zj)=1$ points at the next most
interesting hypothesis for such scenarios. The $Wj$ channels  provide
otherwise sensitive decay modes for the forthcoming searches, under the
requirement of including more data. Finally, no conclusions can be
drawn for scenarios with a single $T$ quark decaying exclusively to Higgs
bosons and jets, as BR$(Hj)=1$ remains a very challenging
hypothesis to probe at the LHC. Dedicated searches might however reach an
impressive sensitivity by relying on EW single production to access
the corresponding topologies (see, e.g., \cite{Carmona:2012jk,Atre:2013ap}).

\subsection{Constraints on realistic scenarios}

As already discussed, in realistic scenarios, the XQs will not be coupled to the third generation quark only, since there are no  stringent theoretical motivations to forbid couplings with SM light quarks. Let us consider this more general situation by using simplified toy models.

As a first toy model, we consider the model of composite top partners presented in \cite{DeSimone:2012fs}, with the only difference that the XQs couple to light generations instead of the third one.
This can be seen as a simplified model for the Kaluza-Klein (KK) resonances associated to the light quarks in models based on warped extra dimensions.
As in the previous Section, we will scan over the masses of the two doublets ${X_{5/3},X_{2/3}}$ and ${T,B}$, assuming that they can vary independently, but keeping the BRs fixed
\begin{eqnarray}
&{\rm BR}(X_{5/3} \rightarrow W^{+}u)={\rm BR}(B\rightarrow W^{+}u)=100\%& \nonumber\\
&\left\{ 
\begin{array}{l}
{\rm BR}(X_{2/3}\rightarrow Zu)={\rm BR}(X_{2/3}\rightarrow Hu)=50\% \\ 
{\rm BR}(T\rightarrow Zu)={\rm BR}(T\rightarrow Hu)=50\%%
\end{array}
\right. \nonumber&
\end{eqnarray}
where $m_{X_{5/3}}=m_{X_{2/3}}$ and $m_{T}=m_{B}$.
The results are shown in Fig. \ref{fig:RattazziLight}: we see that SUSY searches can provide a bound between 500--600 GeV in this scenario and that an interplay between the two doublets increases the bound to 700--800 GeV in the (quasi)degenerate case.

\begin{figure}[h!!]
\centering
\subfigure[]{\epsfig{file=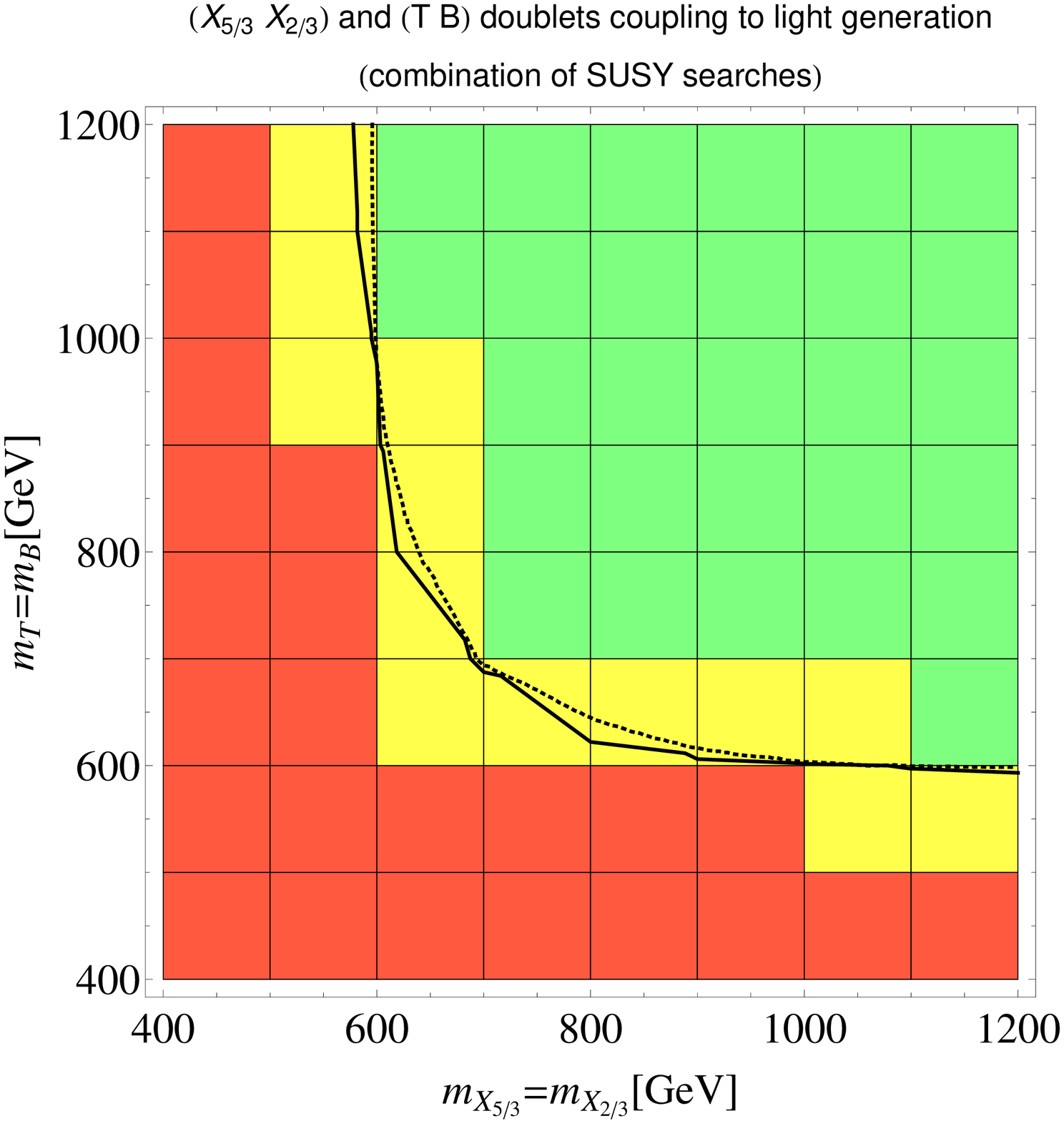,width=.48\textwidth} \label{fig:RattazziLight}}
\subfigure[]{\epsfig{file=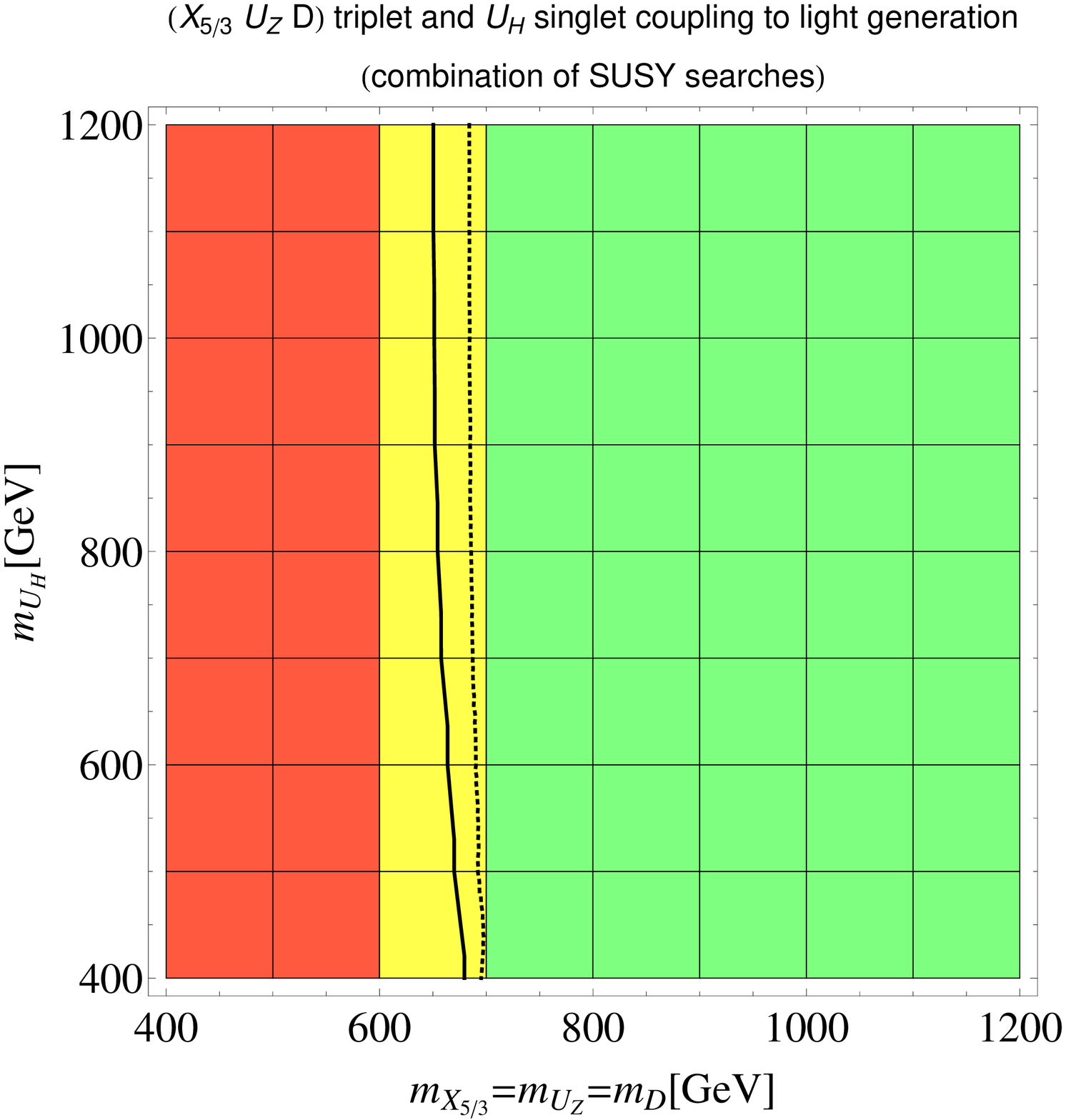,width=.48\textwidth} \label{fig:Panico}}
\caption{\label{fig:multiXQfirst}Excluded (red), boundary (yellow) and non-excluded (green) regions for the $(X_{5/3},X_{2/3})$ and $(T,B)$ doublet model with exclusive couplings to the first generation (a), and for the $(X_{5/3},U_{Z},D)$ and $U_H$ model (b), as presented in \cite{DeSimone:2012fs} and \cite{Atre:2008iu,Atre:2011ae,Atre:2013ap}, respectively. The solid contour corresponds to the 95\% CL boundary obtained by linear interpolation of the eCLs for the simulated points.}
\end{figure}

As a second example, we study the model described in \cite{Atre:2008iu,Atre:2011ae,Atre:2013ap}, also based on a $SO(5)/SO(4)$ CHM.
The particle content is the same as in the previous model \cite{DeSimone:2012fs}, with the difference that the states in the fourplet of SO(4) are  differently rearranged.
In fact, it is assumed that the mechanism that splits the two doublets is suppressed due to the small Yukawa coupling of the light quarks.
What happens instead is that the fourplet decouples into a degenerate
custodial triplet $(X_{5/3},U_{Z},D)$ and a custodial singlet $U_{H}$ (again we keep the notation of the authors for the XQs).
While $X_{5/3}$ and $D$ decay exclusively to $W+j$, the two $Q=2/3$ states
mix, thus giving rise to two distinct physical states. Due
to the custodial symmetry, $U_{Z}$ only couples to $u_{R}^{SM}$
via neutral EW gauge bosons (hence, the $Z$) while $U_{H}$ does so
only through the Higgs boson. In terms of physical eigenstates, a
simplified set of BRs, the ones we used to set bounds, is
\begin{eqnarray}
&&{\rm BR}(X \rightarrow Wj)={\rm BR}(D\rightarrow Wj)={\rm BR}(U_{Z}\rightarrow
Zj)={\rm BR}(U_{H}\rightarrow Hj)=100\% \nonumber
\end{eqnarray}%
We also take the mass splitting $|M_{1}-M_{2}|$  as a free parameter, with $M_{1} =m_{X}=m_{D}=m_{U_{Z}}$ the common mass of the triplet, and 
$M_{2} =m_{U_{H}}$.
The results are shown in Fig.~\ref{fig:Panico}, where we see that SUSY searches set a 95\% CL bound in the 600--700 GeV range on the triplet mass while the mass of the singlet $U_{H}$ is left unconstrained. This is due to the poor sensitivity of the current searches to decays into Higgs plus a light jet.

\section{Conclusions}

Now that the possible existence of a fourth generation of SM-like (i.e., chiral) quarks has essentially been ruled out by the LHC
in the light of the latest Higgs data, an inevitable shift of focus in the search for new extra heavy quarks has been occurring. Prime candidates in playing a center-stage role in this re-newed quest are heavy vector-like quarks, as these arise in a variety of well-motivated BSM scenarios (even if the non-chirality of the new BSM quarks is not a stringent requirement). As a consequence, the more stringent the bounds on such objects are established, the more constrained such new physics models are.

If one looks closely, though, to the phenomenological approaches so far typically exploited in the search for new extra Quarks (XQs), it is immediate to realise that these are rather limiting from the above viewpoint. In fact, XQs generally come numerous in any theoretical BSM scenario, while experimental searches generally only look for one such states. While this approach is well motivated from the experimental point of view, it becomes a non trivial task to translate the bounds in scenarios with multiple resonances. In order to remedy this obvious drawback, we have developed and presented, in this paper,  the \verb|XQCAT| code, a tool which recasts the
available results from the aforementioned experimental searches into those applicable to any spectrum of heavy quark states decaying to $W$, $Z$ and $H$ bosons plus\ jets, top and/or bottom quarks. 
For any implemented
search (or combination thereof), our program \verb|XQCAT| evaluates the
efficiencies and the number of events passing all corresponding cuts. Our
outputs are presented as eCLs on the corresponding
parameter spaces. 

Despite some limitations exist in our approach to multi-XQ quark scenarios, insofar as not accounting for
possible non-SM-like (including cascade) decays, interference effects and some higher order corrections,
\verb|XQCAT| proved robust and flexible enough in starting the pursuit of the above aim. After validating the code
against the inclusive search CMS-B2G-12-015 \cite{Chatrchyan:2013uxa} for a vector-like $T$ quark and, as a novelty,  the indirect SUSY\ searches of 
\cite{Chatrchyan:2012wa}, \cite{Chatrchyan:2012sca}, \cite{Chatrchyan:2012te}%
, \cite{Chatrchyan:2012sa}, \cite{Chatrchyan:2013lya} and \cite%
{Chatrchyan:2012paa} (these are CMS searches only, but this is only a choice, not a limitation of the program),
we provided a re-interpretation on scenarios with one or multiple top
partners, the latter being pair produced via $pp\rightarrow Q\bar{Q} +$ jets,
with the quarks $Q$ having general couplings to the light and third
generations of SM quarks. The {\rm eCL}s were then determined using various methods
beyond the most customary and simplest approach of interpolating linearly the efficiencies.
While conservative, this approach is accurate and fully model-independent. 

The inclusion of non-XQ searches in the data sample exploited by \verb|XQCAT| proved a
crucial ingredient in furthering the scope of the available experimental information, as
we show that the re-interpretation of the SUSY indirect searches (e.g.,
zero-lepton, one-lepton, opposite-sign and and same-sign dilepton signals)
gives robust hints regarding exclusion of XQs below the TeV\ scale at the
LHC. This statement remains true no matter the assumptions on their couplings, yielding a scope comparable and, in some cases, complementary to
the one afforded by XQ direct searches. Further,  their re-interpreted limits  from these and from the
combination of several SUSY searches (i.e., in an approach combining multiple topologies)
appear to be in the same ballpark for pair-produced top partners coupling
dominantly to the third generation.\ 

In short, we have  suggested an analysis framework for models with extra heavy quarks
which we have implemented into the \verb|XQCAT| package. We have sucessfully validated and applied it
to several new physics scenarios and have set up new bounds on models with one or multiple top partners coupling to the SM quarks from
the first generation, third generation or both.
 Limits on the
corresponding parameter spaces have been set from our re-interpretation of
recent direct and indirect searches. We report on the
possibility to set bounds on various benchmark scenarios with non-exclusive
couplings to the third generation. Doing so, we highlight the importance of
combining multiple topology searches to obtain more accurate
(model-independent) re-interpretations of standard single heavy quark searches. 

\bigskip

As an outlook, we would like to highlight the following planned developments of \verb|XQCAT| (not necessarily in chronological order):
\begin{itemize}
\item \underbar{\textit{Inclusion of further experimental searches in the efficiency database:}} currently, the only direct search of vector-like quarks implemented in the tool is \cite{Chatrchyan:2013uxa}, which is designed to identify quarks with charge 2/3, but the implementation of searches dedicated to the observation of quarks with different charges is in order and is currently under development. Furthermore, searches inspired by different BSM scenarios will be included, to test their sensitivity to specific scenarios and compare them with the sensitivity of direct searches. Of course, the most recent results by both ATLAS and CMS will be represented in the tool.
\item \underbar{\textit{Inclusion of EW single production processes:}} single production can be described in a model-independent way, and through the formalism presented in \cite{Buchkremer:2013bha} the signal generated by processes of single production can be decomposed in a limited number of independent subprocesses as in the case of pair production. This upgrade will allow us to provide more stringent constraints, also taking into account experimental searches for single production.
\item \underbar{\textit{Addressing the issues described in Sect.\ref{issues}}}
and closely study  factors which  could affect the conservative estimate of the limits.
\item \underbar{\textit{Implementation of decays into dark matter:}} extra quarks could decay into a SM quark and a new, neutral and stable vector or scalar, which is a Dark Matter candidate. Experimental searches of new quarks decaying to DM candidates have been performed \cite{Aad:2012uu}, and their inclusion in our framework will allow to test these different decay scenarios.
\end{itemize}
Our tool \verb|XQCAT| is currently available at the website \href{https://launchpad.net/xqcat}{https://launchpad.net/xqcat}.

\paragraph{Acknowledgements}
DB, AB, SM and LP  are financed in part through the NExT Institute.
DB, SDC and LP would like to thank the Galileo Galilei Institute (GGI) in Florence for hospitality 
while part of this work was carried out.
The work of MB is supported by the National Fund
for Scientific Research (F.R.S.-FNRS, Belgium) under a FRIA grant. 
AD is partially supported by Institut Universitaire de France. GC and AD also 
acknowledge partial support from Labex-LIO (Lyon Institute of Origins) under grant ANR-10-LABX-66
and from IN2P3 Theory LHC-France funding.

\newpage
\appendix

\section{Determination of the eCL}
\label{app:eCL}

Once  the number of background events (including their uncertainty) is estimated for a given number of observed events, the eCL for a specific signature for a given number of signal events is expressed via ratio of p-values of the Poissonian distributions for the background-only hypothesis  and  signal-plus-background hypothesis as follows\cite{Read:2000ru,Read:2002hq}:
\begin{equation}
 eCL \equiv 1- CL_s = 1-\frac{CL(s+b)}{CL(b)} = 1-\frac{1-\mbox{p-value}(s+b)}{1-\mbox{p-value}(b)}.
\end{equation}
This formula can be extended straightforwardly to the case of multiple channels (or bins of the analysis) by introducing products of $p$-values.

This statistical analysis is not in general as refined as those performed in the experimental studies by the ATLAS and CMS collaborations, but we 
reproduce as accurately as possible the experimental bounds, considering the fact that the raw data are not public. This approximation 
is therefore quite reasonable, as our aim is to obtain new bounds without performing a full analysis, not to optimise the analysis to obtain the best possible bound as typically done in a dedicated search.

The accuracy in the determination of the eCL for a general scenario where the masses of the XQs are different from the simulated 
ones is limited by two factors: firstly, the impossibility of fully reproducing the experimental selection and kinematical cuts;
secondly, the size of the gaps between the simulated masses. 
The first factor can only be quantified by performing validation steps and the full validation procedure is described in Sect.~\ref{sec:validation}. 
Even though the second factor is a technical limitation that can be reduced by performing scans in the XQ masses with smaller mass gaps, one needs to be careful when trying to determine eCLs for generic mass values.
The most accurate results of the tool are obtained when analysing scenarios in which the XQs have masses corresponding to the simulated values: in fact,
in this case, all the factors that contribute to the number of signal events are known. In contrast, if the XQ masses are not corresponding to 
the simulated values, we do not have information about the efficiencies of selection and kinematics cuts. 
Even assuming a smooth behaviour of the efficiency as a function of the mass, there may be non-trivial effects in the mass range, for instance a sudden change in the efficiency when a threshold in the cut of any given bin is passes.
Therefore, one can never be sure to have a correct estimate of the efficiency.
A very fine scan in the masses of the XQs is not possible, due to the computational weight of the MC generation.
There are, however, methods to determine a reliable eCL in the general case. Since none of them alone can provide a completely accurate answer, we adopted different combinations of these methods depending on the specific situation.

\begin{enumerate}

\item \underbar{\textit{Linear interpolation of efficiencies}} The simplest approach is to determine the number of events for a generic mass configuration by linearly interpolating the efficiencies between the closest simulated mass values for each pair-produced XQ in the given scenario. 
This approach assumes that the fluctuations in the efficiencies are small between the simulated values and that the number of signal events is mostly driven by the decrease in the production cross section. This can be a quite strong requirement, especially if the total number of events comes from the interplay of a large number of channels.

\item \underbar{\textit{Determination of a range for confidence levels}} For any given scenario with a number $N$ of XQs, with masses $m_{Q_i}$, $i=1,\dots N$, it is possible to compute the {\rm eCL}s in all the vertices of the $N$-dimensional cube obtained by raising or lowering the input masses to the closest simulated values. 
Assuming that the fluctuations of the efficiencies between the simulated values are not too large and that efficiencies do not drastically increase for increasing masses, the {\rm eCL} of the tested scenario will lie within the minimum and maximum values of the {\rm eCL}s of the hypercube. 

\item \underbar{\textit{Interpolation of {\rm eCL}s}} From the calculation of the {\rm eCL}s in all corners of the $N$-dimensional cube, it is possible to perform an Inverse-Distance-Weighted (IDW) interpolation\cite{IDW} and extract an {\rm eCL} for the input configuration. This approach still assumes that the efficiencies between the simulated mass values have a smooth behaviour.

\end{enumerate}

The first method is by far the less computationally expensive, as the other methods require the calculation of {\rm eCL}s in every corner of an $ N$-dimensional cube, which can be challenging for scenarios with many XQs, as the number of {\rm eCL}s that must be computed scales as $2^N$. Conversely, these more involved methods provide a more accurate, though conservative, determination of the eCL range for a given scenario.
It must be added, however, that an increase in the density of the mass scan for the simulation makes the first approach more and more equivalent to the other approaches. 
Nevertheless, in all of the analyses performed in this preliminary study, we found that the 3 methods give bounds in the mass parameters which are very close to each other and within the intrinsic systematic uncertainties of this approach, like the approximate implementation of detector effects and experimental cuts.

As a last remark about the determination of the eCL, in some specific cases an accurate estimate cannot be performed, and lower bounds on the eCL can only be provided. If the efficiencies increase for increasing mass and if the relative increase in the efficiency values between two simulated masses is comparable (in absolute value) with the relative decrease in the production cross section, the determination of the eCL for a given point within the masses cannot be rigorously done. In this case it can happen that the eCL for points inside the $N$-dimensional cube are either bigger or smaller than all the values in the corners. This is due to the fact that for each point, the number of signal events is computed considering the product of two monotonic functions: the cross-section (decreasing with mass) and the linear interpolation of the efficiencies (which can either increase, decrease or be flat). Scenarios of this kind arise, e.g., when a kinematical cut allows events only above some mass threshold: 
efficiencies that were zero below the threshold, suddenly become sizeable and can determine a sharp increase in the number of signal events. In this case a linear interpolation between efficiencies could overestimate (underestimate) the 
{\rm eCL} for values below (above) the threshold at which the efficiency `switches on'. 
Again, these scenarios will be more and more uncommon as the density of the mass scan increases.

\section{Selected numerical results} \label{app}

\begin{sidewaystable}[t!]
\centering\footnotesize
\begin{tabular}{c|cc|cc|cc}
\toprule
\multirow{3}{*}{Scenario} & \multicolumn{2}{c|}{Range} & \multicolumn{2}{c|}{{\rm eCL}s interpolation} & \multicolumn{2}{c}{efficiencies interpolation} \\
& \multirow{2}{*}{B2G-12-015} & SUSY & \multirow{2}{*}{B2G-12-015} & SUSY & \multirow{2}{*}{B2G-12-015} & SUSY \\
&&combination &&combination && combination \\
\midrule
\midrule
\begin{tabular}{lll} \multicolumn{3}{c}{\textbf{T singlet mixing with first generation only}} \\ BR$(Wb)=0$ & BR$(Zt)=0$ & BR$(Ht)=0$ \\ BR$(Wd)=0.5$ & BR$(Zu)=0.25$ & BR$(Hu)=0.25$ \end{tabular} & $<$400 & 400-500 & \x & 477 & \x & 496  \\
\midrule
\begin{tabular}{lll} \multicolumn{3}{c}{\textbf{T singlet mixing with third generation only}} \\ BR$(Wb)=0.5$ & BR$(Zt)=0.25$ & BR$(Ht)=0.25$ \\ BR$(Wd)=0$ & BR$(Zu)=0$ & BR$(Hu)=0$ \end{tabular} & 600-700 & 500-600 & 643 & 563 & 674 & 590  \\
\midrule
\begin{tabular}{lll} \multicolumn{3}{c}{\textbf{T singlet mixing with first and third generation}} \\ BR$(Wb)=0.25$ & BR$(Zt)=0.125$ & BR$(Ht)=0.125$ \\ BR$(Wd)=0.25$ & BR$(Zu)=0.125$ & BR$(Hu)=0.125$ \end{tabular} & 500-600 & 500-600 & 599 & 518 & 600 & 547 \\
\midrule\midrule
\textbf{T quark decaying 100\% to $W$ + jet} & $<$400 & 400-500 & \x & 417 & \x & 466 \\
\midrule
\textbf{T quark decaying 100\% to $Z$ + jet} & $<$400 & 600-700 & \x & 613 & \x & 636 \\
\midrule
\textbf{T quark decaying 100\% to $H$ + jet} & $<$400 & $<$400 & \x & \x & \x & \x \\
\midrule\midrule
\textbf{T quark decaying 100\% to $W$ + $b$} & 600-700 & 400-500 & 635 & 472 & 672 & 496 \\
\midrule
\textbf{T quark decaying 100\% to $Z$ + top} & 700-800 & 700-800 & 770 & 706 & 789 & 717 \\
\midrule
\textbf{T quark decaying 100\% to $H$ + top} & 600-700 & 600-700 & 630 & 606 & 667 & 614 \\
\midrule\midrule
\textbf{T quark decaying 50\% to $W$ + jet, 50\% to $W$ + $b$} & 600-700 & 400-500 & 607 & 422 & 621 & 473 \\
\midrule
\textbf{T quark decaying 50\% to $Z$ + jet, 50\% to $Z$ + top} & 700-800 & 600-700 & 704 & 632 & 712 & 667 \\
\midrule
\textbf{T quark decaying 50\% to $H$ + jet, 50\% to $H$ + top} & 500-600 & 500-600 & 563 & 516 & 590 & 546 \\
\bottomrule
\end{tabular}
\caption{95\% CL mass bounds in GeV for a pair-produced vector-like quark T, considering representative benchmark scenarios for the branching ratios to first generation quarks, third generation quarks, or both. The explicit numerical values of the limits are obtained through the linear interpolation of the eCLs and of the efficiencies, respectively.}
\label{tab:numericalbounds}
\end{sidewaystable}

In this Appendix we provide numerical results for selected scenarios. The bounds have been obtained by interpolating either the eCLs or the efficiencies between simulated points. We have considered the nominal points with fixed ratios between decays through charged and neutral current or specific scenarios with exclusive decays into one channel, possibly relaxing the assumption of exclusive mixing with one generation. The results are shown in Tab.\ref{tab:numericalbounds}, where all mass values are in GeV.

\clearpage

\bibliography{multivlq}
\bibliographystyle{JHEP}

\end{document}